\documentclass[preprint,12pt]{elsarticle}
\let\today\relax
\makeatletter
\def\ps@pprintTitle{%
    \let\@oddhead\@empty
    \let\@evenhead\@empty
    \def\@oddfoot{\footnotesize\itshape
         {} \hfill\today}%
    \let\@evenfoot\@oddfoot
    }
\makeatother

\usepackage{amssymb}
\usepackage{amsmath}
\usepackage{tabularx}
\usepackage{longtable} % Tables that can span several pages
\usepackage{colortbl}
\usepackage{float}

\definecolor{bluepoli}{cmyk}{0.4,0.1,0,0.4}

\usepackage{booktabs}% http://ctan.org/pkg/booktabs

\begin{document}

\begin{frontmatter}



\title{Hyper-spectral Unmixing algorithms
for remote compositional surface
mapping: a review of the state of
the art} %% Article title

\author[1]{Alfredo Gimenez Zapiola} %% Author name
\author[1]{Andrea Boselli}
\author[1]{Alessandra Menafoglio} %% Author name
\author[1]{Simone Vantini} %% Author name
\affiliation[1]{organization={MOX - Dipartimento di Matematica - Politecnico di Milano},%Department and Organization
            addressline={Piazza Leonardo da Vinci 32}, 
            city={Milan},
            postcode={20133}, 
            country={Italy}}

%% Abstract
\begin{abstract}
%R
%This work provides first a review of the main achievements in hyper-spectral unmixing. Both {linear and nonlinear unmixing} techniques are introduced, that allow to deal with {areal and particulate mixings} of materials respectively. In addition, the most well-renowned public datasets in this setting are shown and described. Finally, some recommendations on where to take future research steps to are provided.

This work concerns a detailed review of {data analysis methods} used for remotely sensed images of large areas of the Earth and of other solid astronomical objects. In detail, it focuses on the problem of inferring the materials that cover the surfaces captured by hyper-spectral images and estimating their abundances and spatial distributions within the region. The most successful and relevant hyper-spectral unmixing methods are reported as well as compared, as an addition to analysing the most recent methodologies. The  most important public data-sets in this setting, which are vastly used in the testing and validation of the former, are also systematically explored. Finally open problems are spotlighted and concrete recommendations for future research are provided. %
\end{abstract}


\begin{highlights}
\item A review of methods, algorithms and dataset is provided for Hyper-spectral Unmixing, the necessary task to provide compositional maps out of hyper-spectral images.
\item The Hyper-spectral Unmixing involves extracting a set of spectral signatures, known as {end members} and their corresponding {fractional abundances} from a hyper-spectral image
\item The standard approach assumes that the mixing of a number of spectra is linear. Their total number and the spectra themselves must be estimated. Several algorithms are available, depending on the chosen mixing models.  Alternative approaches are also explored.

\item Types of datasets in this setting are two-fold: spectral libraries which provide reference spectra, and the hyper-spectral images themselves, which consists of measured reflectances at different wavelengths for each pixel of the image.

\item Open problems are identified, and further research directions are recommended: uncertainty quantification, testing model assumptions statistically and the use of transfer learning.

\end{highlights}

%% Keywords

\end{frontmatter}
\section*{Abbreviations}
HSI, Hyper-Spectral Imaging; HU, Hyper-spectral Unmixing; 



%% Use \subsection commands to start a subsection
\section{Introduction}
\label{sec:intro}
%% Labels are used to cross-reference an item using \ref command.

Remotely sensed images can be an important source of information about vast geographical regions. The interest in them is driven by the large amount of information they store and by their continuous production by airborne and space-borne cameras and sensors.
%Advanced data analysis techniques can be applied on such images to extract valuable information. In particular, {hyper-spectral unmixing} processes hyper-spectral images of geographical regions to (1) infer the materials that cover the regions surfaces together with the respective spectra and (2) determine their spatial distributions. Usually, due to the limited spatial resolution of said images, multiple materials may be contained in the subregion captured by a pixel. Thus, unmixing allows also to estimate {compositional surface maps}, which provide the fractional ground cover of each material in each pixel. 
There exist multiple kinds of such images, depending on the considered spectral bands, the spectral and spatial resolutions and the size of covered area. In particular, {hyper-spectral images} feature a spectrum at each pixel. Each spectrum provides the radiation of the subregion captured by its pixel at multiple narrow wavelength ranges. Typically, hundreds of contiguous ranges in a limited spectral region are considered. 

This review focuses on the problem of identifying the minerals that cover the surfaces present in such hyper-spectral images. It also considers the problem of estimating their abundances and spatial distributions within the region. The two problems are collectively denoted the \textit{hyper-spectral unmixing problem} and have been popular research topics in the last two decades. Particular attention is devoted to the mineralogical applications, where minerals and their spatial distributions are inferred in arid regions with surfacing rocks, both on Earth %especially in arid regions
and other astronomical objects, due to their similarity.

The first objective of this review is to systematically explore and evaluate the advancements of  hyper-spectral unmixing methods and to perform a selection of the most successful ones. The second objective is to gather the most important public data-sets in this setting, since these are those for which the statistical analyses are developed and/or tested to prove their reliability in quality of ideal scenarios where both the materials and their spatial distributions are known. 

Besides providing an updated review including the latest pertinent contributions such as the use of neural networks, this manuscript elaborates on the open problems in the literature, exposes the criticalities to address and provides recommendations for future research directions. 

The text is organised as follows: the following section  supplies the basic concepts regarding hyper-spectral images, the definition and mathematical formulation of the hyper-spectral mixing problem, as well as the main mixing models. Other relevant review articles are also described.  In Section \ref{sec:classical} the main model, i.e., the linear mixing model, is reviewed, providing the necessary mathematical framework; as well as the most popular algorithms that deal with end member extraction and identification. Section \ref{sec:alternative} deals with other methodologies which have been developed more recently and that provide useful alternatives for such classical approach, namely sparse unmixing and nonlinear mixing, the latter of which comprising as an interesting example nonlinear networks. Next, Section \ref{sec:datasets} overviews the best-known datasets for the hyper-spectral unmixing problem, which involves spectral libraries and hyper-spectral cubes as the two types of dataset. 
Relevant and possible further directions of research in the unmixing field and in the analysis of Hyper-spectral cubes are suggested in Section \ref{sec:concl}.

\section{The Unmixing Problem}
\label{sec:Methods_review}%
\subsection{Fundamentals of spectroscopy and hyper-spectral imagery} \label{subsect:spectroscopy}

A key element of hyper-spectral images are of course the {spectra}. The spectrum is the radiation emitted by an object and measured by a sensor at many contiguous narrow ranges in a given spectral region \citet{shaw2003spectral}. The field studying spectra is referred to as {spectroscopy}. In this work, the considered wavelengths range from $0.3$ $\mu m$ to $2.5$ $\mu m$, thus comprising the visible, near-infrared, shortwave infrared regions \citet{bioucas2012hyperspectral}. \\
The upcoming definitions from \citet{shaw2003spectral} better specify the concept of radiation, mentioned above. Foremost, the {reflectance} is the fraction of incident light that is reflected by a surface, while the {reflectivity} is the reflectance spectrum, namely the reflectance as a function of the wavelength. There exist multiple definitions of spectra, like bidirectional or hemispherical or directional, depending on how reflectance is computed \citet{broadwater2010generalized}.

Next, radiance is the power of the light impinging on a surface per unit surface area and unit solid angle of the observation, while {spectral radiance} is the radiance normalised per unit wavelength. It is usually measured in $W/m^2/ \mu m /$steradian. Finally, {single scattering albedo} (SSA) is the “fraction of incoming photons scattered by a particle, divided by the total fraction of photons affected by that particle (either due to scattering or absorption)"  \citet{heylen2014review}. 
In the setting of this review, mostly reflectance spectra or spectral radiance are considered.

{Panchromatic} (or greyscale) {images} feature just one band per pixel, while {RGB images} feature three bands per pixel. {Hyper-spectral images} (HSIs) extend the above including even hundreds of bands, each corresponding to a narrow wavelength range \citet{shaw2003spectral}. 
The images considered in this work are captured on either {airborne} or space-borne hyper-spectral imaging {sensors} \citet{ghamisi2017advances}, and a summary of the main case studies is provided in Section \ref{section:Hyper_cubes}. 
HSIs are usually represented as data cubes, with two spatial dimensions and several spectral dimensions, viz. one for each measured wavelength \citet{bioucas2012hyperspectral}. In particular, each pixel is associated to a reflectance spectrum or a radiance spectrum captured in the corresponding area, and each {plane} stores the reflectance or the radiance captured at a given wavelength in the entire region. 
Reflectance data which characterise a material are more suitable for {material identification} than radiance data, that instead depend on the illumination conditions \citet{shaw2003spectral}. Unfortunately, converting radiance into reflectance is not easy, even if the illumination conditions are known, since it is not trivial to compensate with {atmospheric effects}. An example of a solution for the latter is the model-based atmospheric correction program called ATREM \citet{ATREM}.

\subsection{The Hyper-spectral Unmixing (HU) Problem} \label{subsect:HU_problem}

Each pixel of a HSI may cover a region that includes more than one pure material, due to the limited spatial resolution of the hyper-spectral camera or the strong mixing of materials. In this scenario, {hyper-spectral unmixing} (HU) extracts a set of spectral signatures, known as {end members}, and the corresponding {fractional abundances} from the HSI \citet{bioucas2012hyperspectral}. Each end member should ideally be the spectrum of a pure material included in the image, though the concept of pure material is task-dependent. The fractional abundance of an end member at a given pixel is associated to the portion of ground cover of the corresponding material within that pixel. 
Supposing that a HSI in atmospherically corrected reflectance units is available, the typical approach to HU consists of: (1) estimating the number of end members in the HSI using unsupervised approaches; (2) extracting the spectral signatures of the pure materials from the HSI in the {end member extraction} step; (3) estimating the {abundance maps}, that report the fractional abundances of each end member at each pixel, in the {abundance estimation} step. 
Actually, this general approach admits many variants. For instance, the end members and their number may be a priori known, or a library with hundreds of spectral signatures may be employed, in this case promoting sparsity in the abundances at the abundance estimation step \citet{iordache2011sparse}. Moreover, some methods simultaneously estimate the end member signatures and the abundance maps \citet{miao2007endmember}. In case of high spatial resolution of the HSI and absence of mixing at particle level, {hyper-spectral image classification} can be employed, which associates only one material to each pixel \citet{bioucas2013hyperspectral}.

\subsection{Mixing models} \label{subsect:Mix_models}

As described in \citet{bioucas2012hyperspectral}, mixing within a pixel may be {linear} (additive) or {nonlinear} (beyond additive), depending on the nature of the interaction between the light and the materials, and this distinction further affects the interpretation of the fractional abundances. 
Mixing is linear if the pure materials are segregated in macroscopically distinct regions. It is the poor spatial resolution of the instrument that causes the light to mix within the sensor after impinging on just one material. In this scenario, the fractional abundances represent the fractional areas of the materials within each pixel, and the pixel spectra are regarded as linear combinations of the end member spectra, with the abundances as weights, as detailed in Section \ref{section:LMM}. 
Mixing is nonlinear when light interacts with multiple materials before reaching the sensor, and can be further specified as either {multilayered} or {intimate}. Multilayered mixing occurs instead when light is reflected on multiple surfaces of different pure materials before reaching the sensor. Usually, reflections on more than two materials yield minor effects, so they are neglected, leading to the {bilinear} model. Intimate mixing occurs when pure materials are mixed at microscopic (i.e. particle) level. The most effective method to deal with the latter scenario is based on \textit{Hapke model} \citet{hapke1981bidirectional,heylen2014review,heylen2015multilinear}. The method converts the reflectance spectra into SSA spectra, also relying on simplified formulas, and expresses the pixel SSAs as linear combinations of the SSAs of the pure materials. In this scenario, the fractional abundances represent the mean {cross-sectional areas} of the pure materials.

\subsection{Constraints on unmixing solutions} \label{subsect:constraints_HU}

Many HU algorithms enforce some constraints on the end member signatures or on the fractional abundances \citet{bioucas2012hyperspectral}. First, the extracted end members must be nonnegative. Second, \textit{abundance non-negativity constraint} states that the abundances must be nonnegative. Finally, the \textit{abundance sum constraint} states that the abundances of each pixel must sum to one. The vector of the abundances at a given pixel is called \textit{composition} if these are enforced. In this case, HU provides \textit{compositional surface maps}, that associate a composition to each pixel. 
Actually, while the first two constraints are usually accepted, abundance sum constraint \eqref{eq:ASC} is more criticised, because some materials in the scene might not have been kept into account \citet{bioucas2012hyperspectral} or because the end member signatures may be variable \citet{bioucas2013hyperspectral}, as explained in Subsection \ref{subsect:advances}. Thus, \eqref{eq:ASC} is often discarded, and HU yields estimates of the compositional maps in which this constraint is not fully satisfied.

\subsection{Main references on hyper-spectral unmixing}

Many reviews have been published in the last two decades concerning HSI processing, with a particular focus on HU. This article aims to provide an updated contribution, along the lines of two key publications:  \citet{keshava2002spectral} and \citet{bioucas2012hyperspectral}. \citet{keshava2002spectral} deal with dimensional reduction for HSIs, end member extraction, abundance estimation and distinction between linear and nonlinear mixing models.
\citet{bioucas2012hyperspectral} updates and expands \citet{keshava2002spectral} with a focus on the linear mixing model, introduced in Section \ref{section:LMM}: a taxonomy of linear unmixing methods for end member extraction and an introduction to sparse unmixing are provided. In the present work, novel methods, including sparse unmixing,
nonlinear mixings, statistical modelling, neural networks (cfr. Sections \ref{sec:alternative} and \ref{subsect:MV_methods}) are included, and publicly accessible and vastly utilised data-sets are surveyed.
Concerning nonlinear models, two relevant reviews with a background in signal processing (viz. \citet{heylen2014review} and   \citet{dobigeon2013nonlinear}), have been published. In particular, \citet{heylen2014review} illustrates nonlinear mixtures like bilinear and intimate ones, the Hapke model for unmixing intimate mixtures, and unmixing methods based on neural networks, kernel methods and support vector machines. Also, manifold learning techniques and piecewise linear methods are considered. 
Finally, \citet{bioucas2013hyperspectral} provides a broader picture of hyper-spectral remote sensing, as it deals with unmixing and classification, but also with target detection, physical parameter retrieval and fast computing. \citet{ghamisi2017advances} further updates \citet{bioucas2013hyperspectral}.

\section{Classical methodologies}
\label{sec:classical}

\begin{figure}[t]%% placement specifier
%% Use \includegraphics command to insert graphic files. Place graphics files in 
%% working directory.
\centering%% For centre alignment of image.
\includegraphics[width=80mm]{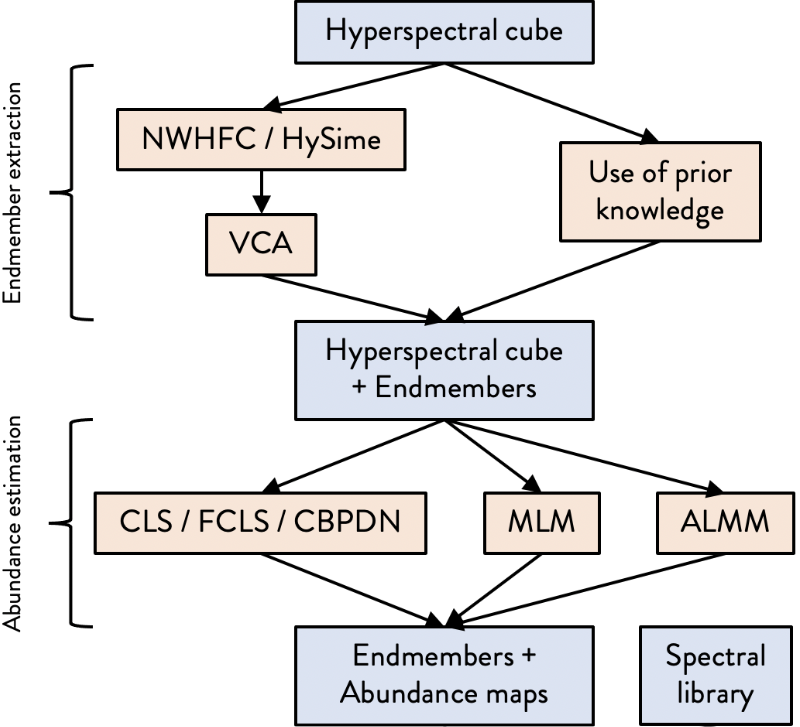}
%% Use \caption command for figure caption and label.
\caption{Workflow for solving the HU problem: classical methodology}\label{fig1}
%% https://en.wikibooks.org/wiki/LaTeX/Importing_Graphics#Importing_external_graphics
\end{figure}

In this Section, the methods that have been utilised the most since the beginning of the HU field are outlined. Firstly, the mathematical models that represent the abundance of several minerals in the pixels of the Hyper-spectral cubes are explained (Subsections \ref{section:LMM} and \ref{subsect:MLM}). Whereas these could be known a priori, that is seldom the case, and hence the estimation of the number of end members and their extraction are also part of solving the HU problem, and are outlined in Subsections \ref{subsec:number_end members} and \ref{subsec:end member_extraction}. The overall workflow is summarised in Figure \ref{fig1}.

\subsection{Linear mixing model} \label{section:LMM}

The linear mixing model is the most popular mixing model in the literature, especially for areal mixtures. Mathematically, it is constructed in a setting of linear combinations of vectors constrained to be positive and summing to one, namely the convex geometry.
The following are some basic notions on said subject, based on  \citet{ambikapathi2011chance,chan2009convex,chan2011simplex, bioucas2012hyperspectral}. Consider a set of $p$ vectors $\{\mathbf{v}_k \}_{k=1}^p$ in $\mathbb{R}^B$. Their {convex hull} is defined as
\begin{equation}
    \text{conv}\{ \mathbf{v}_1, ..., \mathbf{v}_p \} = \left\{ \sum_{k=1}^p \theta_k \mathbf{v}_k \quad \text{s.t.} \quad \sum_{k=1}^p{\theta_k} = 1, \quad \theta_k \geq 0 \quad \forall k \right\}.
\end{equation}
\noindent A {vertex} of $\text{conv}\{ \mathbf{v}_1, ..., \mathbf{v}_p \}$ is an extremal point of the hull. The $\text{conv}\{ \mathbf{v}_1, ..., \mathbf{v}_p \}$ is denoted $(p-1)$-\textit{simplex} if  $\{\mathbf{v}_k \}_{k=1}^p$ are {affinely independent}, namely if $\{ \mathbf{v}_k - \mathbf{v}_1 \}_{k=2}^p$ are linearly independent. In this case, its vertices coincide with $\{\mathbf{v}_k \}_{k=1}^p$. \\
Finally, the {affine hull} of $\{\mathbf{v}_k \}_{k=1}^p$ is defined as
\begin{equation}
    \text{aff}\{ \mathbf{v}_1, ..., \mathbf{v}_p \} = \left\{ \sum_{k=1}^p \theta_k \mathbf{v}_k \quad \text{s.t.} \quad \sum_{k=1}^p{\theta_k} = 1 \right\}.
\end{equation}

An HSI with $n$ pixels and $B$ spectral bands can be represented as $\mathbf{Y} = [y_{ij}]_{i,j} = [\mathbf{y}_1, ..., \mathbf{y}_n]$ in $\mathbb{R}^{B,n}$, where $y_{ij}$ is the spectrum at band $i$ of pixel $j$ and, thus, $\mathbf{y}_j$ is the spectrum vector at pixel $j$. The {linear mixing model} (LMM) approximates each pixel spectrum $\mathbf{y}_j$ with a linear combination of end member spectra, with the corresponding fractional abundances as weights \citet{bioucas2012hyperspectral}:
\begin{equation} \label{eq:LMM_scalar}
    y_{ij} = \sum_{k=1}^p m_{ik} x_{kj} + w_{ij}, \quad i = 1,...,B \quad j = 1,...,n
\end{equation}
\noindent where $p$ is the number of end members in the scene, $m_{ik}$ is the spectrum at band $i$ of end member $k$, $x_{kj}$ is the fractional abundance of end member $k$ at pixel $j$ and $w_{ij}$ is an additive noise or modelling error at band $i$ of pixel $j$. Denoting by $\mathbf{m}_k = [m_{1k},...,m_{Bk}]^T$ the spectrum of end member $k$, $\mathbf{M} = [\mathbf{m}_1,...,\mathbf{m}_p]$ the \textit{mixing matrix}, $\mathbf{w}_j = [w_{1j},...,w_{Bj}]^T$ the noise spectrum at pixel $j$, and $\mathbf{x}_j = [x_{1j},...,x_{pj}]^T$ the abundance vector at pixel $j$, equation \eqref{eq:LMM_scalar} can be cast in vector form
\begin{equation} \label{eq:LMM_vector}
    \mathbf{y}_j = \mathbf{M} \mathbf{x}_{j} + \mathbf{w}_{j}, \quad j = 1,...,n.
\end{equation}
\noindent Thus, given a \textit{data matrix} $\mathbf{Y}$, end member extraction estimates the mixing matrix $\mathbf{M}$ (including the number of end members $p$), while abundance estimation retrieves the abundance vector $\mathbf{x}_j$  for each pixel $j$, relying on equation \eqref{eq:LMM_scalar}. Employing matrices $\mathbf{X} = [\mathbf{x}_1,...,\mathbf{x}_n]$ and $\mathbf{W} = [\mathbf{w}_1,...,\mathbf{w}_n]$, that gather the abundance vectors and the noise vectors for all pixels respectively, equation \eqref{eq:LMM_scalar} can also be cast in matrix form as
\begin{equation} \label{eq:LMM_matrix}
    \mathbf{Y} = \mathbf{MX} + \mathbf{W}.
\end{equation}
Using the above notation, abundance non-negativity constraint and abundance sum constraint assumptions (see Subsec. \ref{subsect:Mix_models}) are respectively cast as:
\begin{equation*}  \label{eq:ANC}
    x_{kj} \ge 0, \quad k = 1,...,p, \quad j = 1,...,n
\end{equation*}
\begin{equation*} \label{eq:ASC}
    \sum_{k=1}^{p}x_{kj} = 1, \quad j = 1,...,n.
\end{equation*}
\noindent Enforcing both, all the abundance vectors $\{\mathbf{x}_j\}_{j=1}^n$ lay in a $(p-1)$-simplex. Moreover, if the columns of $\mathbf{M}$ are affinely independent, their convex hull is a $(p-1)$-simplex in $\mathbb{R}^B$. Most algorithms based on LMM perform end member extraction by retrieving the vertices of such simplex. \\
Once the matrix $\mathbf{M}$ has been extracted, a typical approach to retrieve the abundances $\mathbf{x}_j$ at each pixel $j$ is to compute the \textit{least squares} (LS) solution
\begin{equation} \label{eq:LMM_LS}
    \min_{\mathbf{x}_j} \|\mathbf{y}_j - \mathbf{M} \mathbf{x}_j\|_2
\end{equation}
\noindent where $\|\cdot\|_q$ is the $q$-norm in $\mathbb{R}^B$ \citet{iordache2011sparse}. Imposing \eqref{eq:ANC} on \eqref{eq:LMM_LS} yields the \textit{nonnegative constrained least squares} (CLS) solution, while imposing both \eqref{eq:ANC} and \eqref{eq:ASC} yields the \textit{fully constrained least squares} (FCLS) solution \citet{iordache2011sparse}.

\subsubsection{Bilinear models}

As mentioned in Section \ref{subsect:Mix_models}, mixing is bilinear when light is assumed to interact with two materials before reaching the sensor \citet{heylen2014review}. The resulting interaction between the respective materials, with spectra $\mathbf{m}_k$ and $\mathbf{m}_h$, is modelled through a new spectrum $\mathbf{m}_k \odot \mathbf{m}_h$, consisting of the bandwise product of $\mathbf{m}_k$ and $\mathbf{m}_h$. As the bandwise product of the reflectances of more than two end member spectra yields very small reflectances, higher-order interactions are usually negligible. 
With the notation introduced in Section \ref{subsect:Mix_models} and dropping the pixel index $j$ for simplicity, bilinear models express each pixel spectrum $\mathbf{y}$ as
\begin{equation}
    \mathbf{y} = \sum_{k=1}^{p} x_{k} \mathbf{m}_k + \sum_{k=1}^{p}\sum_{h=1}^{p} x_{kh} \mathbf{m}_k \odot \mathbf{m}_h + \mathbf{w}
\end{equation}
\noindent where $\{x_{k}\}_{k=1}^p$ represent the abundances of the pure materials in a given pixel, while the properties and the interpretation of $\{x_{kh}\}_{k,h}$ are model-dependent. The most relevant differences among the models are whether the abundance non-negativity and sum constraints are imposed only to $\{x_{k}\}_{k=1}^p$ or to all the parameters and whether the self-interaction terms $\{x_{kk}\}_{k=1}^p$ are non-zero. 
The major drawbacks of bilinear models are the small magnitude of the interaction spectra, that may be confused with spectra caused by shadows, and the huge increase of the number of parameters to be estimated, that may cause \textit{overfitting}. 
The most relevant bilinear model is called \textit{generalised bilinear model}  \citet{halimi2011nonlinear}. It sets $x_{kh} = \gamma_{kh} x_k x_h$, with $\{x_{k}\}_{k=1}^p$ satisfying \eqref{eq:ANC} and \eqref{eq:ASC} conditions, while $\gamma_{kh} = 0$ if $k \geq h$ and $\gamma_{kh} \in [0,1]$ otherwise. Parameters are inferred using hierarchical Bayesian methods, featuring MCMC techniques.

\subsection{Multilinear model} \label{subsect:MLM}

The {multilinear model}  \citet{heylen2015multilinear} extends the bilinear model, introducing infinite interactions. This allows to cope with different mixing schemata, including linear and intimate ones, and in particular with mineral mixtures, where multiple reflections are more likely to happen. With respect to classic LMM, it introduces just one additional parameter $P$, resulting in a simple model without risk of overfitting. 
Light is modelled as a ray that moves along different material particles as a \textit{Markov chain}. $P$ represents the probability of interacting with a new particle, while each abundance $x_j$ is proportional to the probability of hitting a particle of material $j$. Referring to the notation of Subsection \ref{section:LMM}, dropping the pixel index $j$ for simplicity, and neglecting the noise, the pixel reflectance $\mathbf{y}$ is the sum over every existing path, that results to be
\begin{equation}
    \mathbf{y} = \frac{(1-P)\sum_{k=1}^p x_k \mathbf{w}_k}{1-P\sum_{k=1}^p x_k \mathbf{w}_k}
\end{equation}
\noindent where $\{ \mathbf{w}_k \}_{k=1}^p$ are the SSA spectra of the end members, supposed known, and the division is performed component-wise. Actually, in remote sensing applications, small values of $P$ can be assumed, and $\mathbf{w}_k$ can be replaced with the corresponding reflectance $\mathbf{m}_k$. The quantities
$\mathbf{x}$ and $P$ for each pixel are simultaneously retrieved minimizing the reconstruction error of the pixel spectrum $\mathbf{y}$
\begin{equation} \label{eq:MLM_equation}
    (\hat{\mathbf{x}}, \hat{P}) = \text{arg} \min_{\mathbf{x},P} \left\| \mathbf{y} - \frac{(1-P)\sum_{k=1}^p x_k \mathbf{m}_k}{1-P\sum_{k=1}^p x_k \mathbf{m}_k} \right\|_2^2
\end{equation}
\noindent enforcing non-negativity and sum constraints (\eqref{eq:ANC} and \eqref{eq:ASC}) for $\mathbf{x}$ and $P < 1$. 
The value of $P$ provides the level of nonlinearity at each pixel and the model is physically meaningful if $P \in [0,1)$, being $P$ a probability. Actually, also negative values of $P$ can be accepted, as they are associated to an increase in the pixel reflectance with respect to the one obtained with LMM.

\subsection{Estimation of the number of end members}\label{subsec:number_end members}

Most end member extraction methods require the number of end members $p$ to be a priori known. If $p$ is not known, it can be estimated through suitable techniques, that are roughly divided in three classes in \citet{tao2021endmember}. \\
The first class consists of {information theory-based algorithms}, that employ criteria such as Akaike's Information Criterion, Bayesian Information Criterion  and Minimum Description Length  for model selection. Some of these methods may overestimate $p$ on real data \citet{chang2004estimation}. 
A second, small class consists of {geometry characterisation algorithms}. In particular, at each iteration, the algorithms proposed by \citet{ambikapathi2012hyperspectral} estimate a new end member spectrum using a {greedy} end member extraction algorithm, until a convex or affine hull with maximum volume is achieved, respectively. 
The third and most relevant class includes the {eigenvalue thresholding methods}, in which the estimate of $p$ depends on the result of an eigen-analysis of the pixel spectra. The most representative methods in this class are {principal component analysis}, {noise-whitened Harsanyi-Farrand-Chang}  \citet{chang2004estimation}, {eigenvalue likelihood maximization} \citet{luo2012empirical} and {hyper-spectral signal subspace by minimum error} \citet{bioucas2008hyperspectral}.

%\subsubsection{NWHFC and ELM algorithms} %\label{subsect:NWHFC}

%\subsubsection{HySime algorithm} %\label{subsect:HySime}

Another very popular approach for estimating $p$ is HySime \citet{bioucas2008hyperspectral}, which belongs to the eigenvalue thresholding methods and does not depend on any parameter. For related approaches, see \citet{chang2004estimation} and \citet{luo2012empirical}.
First, the noise is estimated from the data, which is used to approximate the signal and compute the signal and the noise sample correlation matrices. Next, a subspace spanned by $p$ eigenvectors of the sample signal correlation matrix is retrieved. This is done minimising the {mean squared error} (MSE) between the estimated signals and the projections in the subspace of the corresponding noisy spectra. Minimization is performed with respect to $p$, yielding the chosen subset eigenvectors.

%\subsubsection{Geometrical pure-pixel algorithms}

\subsection{End member extraction with pure-pixel algorithms}\label{subsec:end member_extraction}

Under the linear mixing model, the other key aspect is the availability of end members that are to be (convexly) combined. Usually, these must be obtained from data. A diffuse approach is the one taken in geometrical pure-pixel algorithms. These are based on the \textit{pure-pixel hypothesis} \citet{bioucas2012hyperspectral}: for each material in the scene, there exists at least one pixel containing that material only. These algorithms retrieve the data simplex selecting its vertices among the available pixel spectra. They are computationally light, but the pure-pixel hypothesis is strong, and may not hold in case of low spatial resolution or intimate mixtures.  
The most employed end member extraction algorithm based on pure-pixel hypothesis --and more in general on LMM-- is {vertex component analysis} (VCA) \citet{nascimento2005vertex}, unfolded later in this subsection. \\
Some pure-pixel methods are based on \textit{Winter's belief}, according to which the simplex having the pure pixel spectra as vertices has the highest volume with respect to any other combination of pixels. The most representative method of this class is N-FINDR \citet{winter1999n}, that progressively grows a simplex within the data cloud. A drawback of N-FINDR is that it is randomly initialised \citet{chang2006new}. 
Some methods are based on N-FINDR. For instance, the \textit{simplex growing algorithm} (SGA) \citet{chang2006new} finds first the 2-simplex with maximum volume and at each step it adds an end member that maximises the q-simplex volume. Differently from N-FINDR, SGA is greedy, so it is computationally lighter. Finally, \textit{alternating volume maximization} \citet{chan2011simplex} performs the simplex volume maximization by changing just one end member per iteration.
Notice that the actual volume of a $(p-1)$-simplex in $\mathbb{R}^B$ is null, so that the volumes mentioned above are meant as quantities computed through specific procedures, an instance of which is detailed in Subsection \ref{subsect:MV_methods}. 
There exist other successful pure-pixel methods, that are not inspired by VCA or N-FINDR. The most relevant is \textit{pixel purity index} (PPI) \citet{boardman1995mapping}, included also in \textit{Environment for Visualizing Images} (ENVI) software \citet{berman2004ice}. All pixel spectra are projected into a huge number of random vectors, and the pixels whose spectra are extremal in most projections are selected as the purest. 

\subsubsection{VCA algorithm} \label{subsect:VCA_algo}
VCA \citet{nascimento2005vertex} exploits the identification of the end members with the vertices of a $(p-1)$-simplex and it searches for such vertices in the set of pixel spectra, relying on the pure-pixel hypothesis. Its performance is comparable to the one of N-FINDR, but it has a much lower computational complexity. \\
VCA is based on an extension of LMM equation \eqref{eq:LMM_vector}
\begin{equation} 
    \mathbf{y}_j = \mathbf{M} (\gamma_j \mathbf{x}_{j}) + \mathbf{w}_{j}, \quad j = 1,...,n.
\end{equation}
where $\gamma_j \in [0,\infty)$ is a \textit{scaling factor} accounting for variable illumination. Thus, neglecting the noise, the pixel spectra belong to the \textit{convex cone}
\begin{equation}
    \mathcal{C}_{p} = \left\{ \sum_{k=1}^p \gamma \theta_k \mathbf{m}_k \quad \text{s.t.} \quad \sum_{k=1}^p{\theta_k} = 1, \quad \theta_k \geq 0 \quad \forall k, \quad \gamma \geq 0 \right\}.
\end{equation}
The first step of VCA is dimensional reduction, that projects the data into the $(p-1)$-dimensional affine subspace containing the $(p-1)$-simplex with the end members as vertices. PCA or \textit{singular value decomposition} (SVD) are performed, depending on the estimated \textit{signal-to-noise ratio} of the data. After projection, at each iteration, the data are projected into a direction orthogonal to the already detected end members, and the extremal pixel spectrum is associated to a new end member. 
The idea underlying VCA is similar to the one of {orthogonal subspace projection}  detection method \citet{harsanyi1994hyperspectral}. It detects a target end member within a pixel by projecting the pixel spectrum first into an orthogonal subspace to some discarded spectra and then into the target end member. OSP requires all the spectra to be a priori known. 
Finally, some methods are derived from VCA, notably {successive volume maximisation}  \citet{chan2011simplex}.

\subsubsection{Geometrical minimum-volume algorithms} \label{subsect:MV_methods}

{Geometrical Minimum-volume} methods look for the simplex with minimum volume containing the pixel spectra \citet{bioucas2012hyperspectral}. To yield good estimates of the simplex, these methods need the presence of at least $(p-1)$ pixel spectra for each simplex facet \citet{iordache2011sparse}. However, they do not require the pure-pixel assumption, as the extracted end member spectra may not belong to the set of pixel spectra. A notable drawback is their higher computational burden with respect to the pure-pixel methods. 

According to \citet{bioucas2012hyperspectral}, the volume mentioned above is obtained by projecting the pixel spectra in a $p$-dimensional subspace $\mathcal{S}$, determined through {dimensional reduction techniques}. Being $\mathbf{P}_{\mathcal{S}}$ the matrix with an orthonormal basis of $\mathcal{S}$ as columns, LMM holds in subspace $\mathcal{S}$ in the form:
\begin{equation}
    \mathbf{y}_{\mathcal{S}} = \mathbf{M}_{\mathcal{S}} \mathbf{x} + \mathbf{w}_{\mathcal{S}}
\end{equation}
\noindent where $\mathbf{y}_{\mathcal{S}} = \mathbf{P}_{\mathcal{S}}^T \mathbf{y}$, $\mathbf{M}_{\mathcal{S}} = \mathbf{P}_{\mathcal{S}}^T \mathbf{M}$, $\mathbf{w}_{\mathcal{S}} = \mathbf{P}_{\mathcal{S}}^T \mathbf{w}$ are the projections into the signal subspace of pixel spectrum, mixing matrix and noise spectrum respectively. Assuming that the columns of $\mathbf{M}_{\mathcal{S}}$ are affinely independent, their convex hull is a $(p-1)$-simplex as in the original space. A non-zero volume in $\mathbb{R}^p$ can be achieved extending the simplex to the origin, though some alternatives are available. 
The most employed method in the minimum-volume class is the \textit{minimum volume constrained nonnegative matrix factorisation} \citet{miao2007endmember}. It simultaneously determines the fractional abundances $\{\mathbf{x}_j \}_{j=1}^n$ and the end member spectra $\mathbf{M}$ in the original spectral space by solving an optimisation problem which involves a two-terms objective function. The first term is the {reconstruction error} of the pixel spectra through the LMM \eqref{eq:LMM_vector}, while the second is related to the volume of the simplex of the end member spectra. The method is computationally demanding, as it exploits all the pixels in the image. 
A popular variant is $L_{1/2} \text{-NMF}$ \citet{qian2011hyperspectral}, that also solves a \textit{nonnegative matrix factorisation} problem, where a sparsity enforcing term is used in place of the simplex volume term. $L_{1/2} \text{-NMF}$  has been broadly studied in the HU field in the last decade, and many models with different constraints and regularisation terms have been introduced \citet{yao2019nonconvex}. 

Another relevant minimum-volume method is \textit{simplex identification via split augmented Lagrangian} \citet{bioucas2009variable}, which achieves robustness with respect to noise employing {soft positivity constraints} on the abundances. Finally, a very similar algorithm is, \textit{minimum volume simplex analysis} \citet{li2015minimum}.

\subsubsection{Statistical algorithms}

If the pure-pixel assumption is not satisfied and there are not at least $(p-1)$ pixel spectra per simplex facet, both pure-pixel and minimum-volume methods fail. Statistical methods can cope with highly mixed scenarios, despite being more computationally intensive \citet{bioucas2012hyperspectral}. 
Most statistical methods are {Bayesian}, where a {prior distribution} is assigned to the mixing matrix $\mathbf{M}$ and to the fractional abundances $ \mathbf{X}$, a {likelihood} is assigned to the data matrix $\mathbf{Y}$, and the {posterior distribution} is computed using the {Bayes formula}. Suitable priors allow enforcing the non-negativity of spectra and abundances and other prior knowledge on the problem.
The most representative among the Bayesian methods is \citet{dobigeon2009joint}, which works in a projected subspace $\mathcal{S}$. It employs a {hierarchical Bayesian model}, where conjugate priors are used for end members and abundances parameters, while non-informative priors are assigned to the hyperparameters of the parameters priors. The posterior is approximated through {Markov-Chain Monte Carlo (MCMC) techniques}, in particular a {Gibbs sampler}, and the {posterior means} of the parameters are estimated. 
Bayesian methods have been employed also for \textit{piecewise linear mixing models} \citet{bioucas2012hyperspectral}: assuming that different regions of the HSI include different sets of pure materials, these models retrieve a simplex for each of these regions. A notable example is \textit{piecewise convex end member detection} \citet{zare2010pce}. 

\subsubsection{End member extraction: alternative algorithms}
\label{subsec:end member-no-pixel}
Whereas most HU methods treat HSIs as a collection of pixel spectra without exploiting any {spatial information}, a few methods consider such information, and they can be divided in two classes \citet{xu2018regional}. 
The first includes algorithms that are specifically designed to consider spatial information. The most notable example is the \textit{automatic morphological end member extraction} \citet{plaza2002spatial}, which allows for both end member extraction and pixel-purity evaluation using {mathematical morphology operators}. Another notable method is \textit{spatial spectral end member extraction} \citet{rogge2007integration}. 
The second class includes {pre-processing methods}, that are applied before any other end member extraction technique to account for the spatial information. They usually select the most suitable pixels to be fed to such end member extraction technique. For instance, {spatial preprocessing}  \citet{zortea2009spatial} assumes that spectrally homogeneous areas are more likely to include useful pixel spectra for end member extraction. Thus, a {pixel similarity metric} is introduced, which promotes the search of end members in such areas.

\subsubsection{Classical methodologies: comparative summary}

An important question arises: which classical workflow should be chosen? For such purpose, two comparative and summarising tables are displayed below. Regarding the type of mixing, possibilities are linear, bilinear or multilinear. The advantage of the first is, apart from its interpretability, that constrained or fully constrained least squares methods are employed for a quick (computationally) model fitting, without the risk for over-fitting. Were interactions to be taken into account, one could either choose which ones to model in the bilinear schema, or else allow for potentially infinitely with the multilinear. One the one hand, the first would require careful prior elicitation, as model fitting is performed through bayesian methods such as MCMC, yet allowing at the same time the possibility to mitigate over-fitting with such chosen priors. The multilinear model is definitely more flexible, yet at the expense of heavy computational burden in solving the reconstruction error (\ref{eq:MLM_equation}), and potentially leading to models which do not generalise well due to over-specification. See Table \ref{tab:linear} for such comparison.

\begin{table}[h!]
    \centering
    \begin{tabular}{cccc}
         Mixing & Fitting & Computational burden & Model complexity \\
         \hline
          Linear&  CLS or FCLS & Low & Low  \\
          Bilinear & Bayesian MCMC  & Low to mild & Mild\\ 
          Multilinear & Error reconstruction  & Very high & High\\
    \end{tabular}
    \caption{Classical methodology comparison}
    \label{tab:linear}
\end{table}

The other modelling choice are the endmember selectors. VCA and Geometric Min-Volume algorithms, as seen above, are inspired directly on the linear mixing assumption, which makes them  natural candidates in such setting. The choice of one over the other will naturally be case-specific. Yet, statistical algorithms as well as alternative endmember selectors exist: should they be used? If the mixing schema goes beyond the linear one, their consideration is direct,
and what is more they may take into account the spatial dependence between contiguous pixels, leading to better solutions to the HU problem. See Table \ref{tab:endmember_extraction} for a summary of this comparison.

\begin{table}[h!]
    \centering
    \begin{tabular}{ccc}
         Endmember selection & Comp. burden & Models spatial dependence\\
         \hline
          VCA& Low  & No\\
          Geometric min-volume & High & No \\
          Statistical & Model-dependent & Yes \\
          Alternative models & Mild to high &  Yes
    \end{tabular}
    \caption{Endbember extraction comparison}
    \label{tab:endmember_extraction}
\end{table}

\section{Alternative  methodologies}
\label{sec:alternative}
\subsection{Sparse unmixing}

%\subsubsection{Sparse unmixing problem} 
\label{subsect:sparse_unmixing_problem}

{Sparse unmixing}, described in \citet{iordache2011sparse, bioucas2012hyperspectral}, provides an alternative workflow to the classical one detailed in section \ref{subsect:HU_problem}. It assumes that the pixel spectra are a linear combination of a few spectra from a large \textit{spectral library}, namely a database of spectra, and it selects the ones that best approximate each pixel spectrum through LMM equation \eqref{eq:LMM_vector}. It is denoted sparse because the number of end members in a scene is usually quite small, so that only few optimal spectra need to be selected from the library. 
The main benefit of sparse unmixing is that the end member extraction step is replaced by the use of a spectral library, so that only the abundance estimation step needs to be performed. A major drawback is that spectra from libraries and from HSIs are usually acquired in different conditions, thus some preprocessing on the pixel spectra often needs to be performed. Another drawback is that some spectra from libraries are usually similar. In particular, libraries show a very high \textit{mutual coherence}, namely the maximal cosine of the angle between any two spectral vectors from the library \citet{iordache2012total}. This complicates the search of sparse solutions for the abundance estimation problem. 

Referring to the notation of the previous section, let $\mathbf{M}$ include $u$ spectra with $B$ bands from a library, where usually $B < u$. Neglecting at first abundance non-negativity / sum constraints (cfr. \eqref{eq:ANC} and \eqref{eq:ASC}), the problem of approximating a pixel spectrum $\mathbf{y}_j$ employing a minimum number of end members is
\begin{equation} \label{eq:SU_P0delta}
    \min_{\mathbf{x}_j}\|\mathbf{x}_j\|_0 \quad \text{s.t.} \quad \|\mathbf{y}_j-\mathbf{M}\mathbf{x}_j\|_2 \leq \delta
\end{equation}
\noindent where $\|\cdot\|_0$ is the number of non-zero elements, and $\delta > 0$ is a threshold on the approximation error. Being \eqref{eq:SU_P0delta} \textit{NP-hard}, it is usually replaced with
\begin{equation} \label{eq:SU_P1delta}
    \min_{\mathbf{x}_j}\|\mathbf{x}_j\|_1 \quad \text{s.t.} \quad \|\mathbf{y}_j-\mathbf{M}\mathbf{x}_j\|_2 \leq \delta.
\end{equation}
It can be shown that an equivalent formulation of \eqref{eq:SU_P1delta} is 
\begin{equation} \label{eq:SU_SR}
\min_{\mathbf{x}_j} \frac{1}{2}\|\mathbf{y}_j-\mathbf{M}\mathbf{x}_j\|_2^2+\lambda\|\mathbf{x}_j\|_1
\end{equation}
\noindent where $\lambda > 0$ is a \textit{Lagrange multiplier}. Enforcing also the non-negativity constraints, problem (\eqref{eq:SU_P1delta}, \eqref{eq:SU_SR}) is denoted \textit{constrained sparse regression}  \citet{bioucas2010alternating,iordache2011sparse,bioucas2012hyperspectral} or \textit{constrained basis pursuit denoising}.  The abundance sum constraint \eqref{eq:ASC} is usually not enforced, because this would imply $\|\mathbf{x}_j\|_1 = 1$, thus yielding an equivalent objective function for FCLS problem.

%\subsubsection{SUnSAL algorithm} \label{subsect:SUnSAL_algo}

{Sparse unmixing by splitting and augmented Lagrangian} (SUnSAL) algorithm, detailed in \citet{bioucas2010alternating}, solves problem \eqref{eq:SU_SR} for $\lambda \geq 0$. It is based on the \textit{alternating direction method of multipliers}, that performs variable splitting and solves a constrained problem through the augmented Lagrangian method. 
Minor modifications to the algorithm allow enforcing both abundance non-negativity \eqref{eq:ANC} and abundance sum \eqref{eq:ASC} constraints, so that SUnSAL is able to solve many important problems: \textit{constrained least squares} (CLS) is solved setting $\lambda = 0$ and enforcing \eqref{eq:ANC}; \textit{fully constrained least squares} (FCLS) is solved additionally enforcing \eqref{eq:ASC}. \textit{Constrained basis pursuit denoising} is solved by choosing a positive $\lambda$ and enforcing \eqref{eq:ANC}. 
The complexity per iteration is $\mathcal{O}(u^2)$, so it heavily depends on the cardinality of the spectral library. As each pixel is processed independently, \textit{parallel computing} techniques can also be employed to speed up the unmixing \citet{iordache2011sparse}.

\subsubsection{Improvements on SUnSAL}

Two very successful improvements on SUnSAL algorithm have been proposed, each coping with some weaknesses of the method, in particular with the high mutual coherence of the spectral libraries. The first is \textit{sparse unmixing via variable splitting augmented Lagrangian and total variation}, that accounts for the spatial information of the HSI \citet{iordache2012total}. In particular, it introduces the {total variation} regularisation term in the objective function, which promotes spatial homogeneity, so that the abundance of each end member varies regularly in nearby pixels. 
With the notation introduced in the previous section, let $\|\cdot\|_F$ denote the {Frobenius norm}, $\lambda$ and $\lambda_{TV}$ be two nonnegative parameters and $\epsilon$ be the set of $\{i,j\}$ indices of two horizontally or vertically neighbouring pixels. Moreover, let
\begin{equation*}
    \text{TV}(\mathbf{X}) = \sum_{\{i,j\} \in \epsilon} \|\mathbf{x}_i - \mathbf{x}_j\|_1
\end{equation*}
\noindent be the TV regularisation term. Then the following optimisation problem is solved:
\begin{equation*}
    \min_{\mathbf{X}} \frac{1}{2}\|\mathbf{Y}-\mathbf{MX}\|_F^2 + \lambda \sum_{i=1}^n\|\mathbf{x}_i\|_1 + \lambda_{TV} \text{TV}(\mathbf{X}) \quad \text{s.t.} \quad x_{kj} \geq 0 \quad \forall k,j.
\end{equation*}
\noindent Notice that constraint \eqref{eq:ASC} is not enforced, and that setting $\lambda_{TV} = 0$ yields the same objective function of \eqref{eq:SU_SR}, as optimisation can be performed pixel-wise. It is again based on the \textit{alternating direction method of multipliers} method.

The second improvement is \textit{collaborative SUnSAL} \citet{iordache2013collaborative}. It assumes that all the pixels are occupied by a small number of end members, equal for all pixels. Consequently, it limits the number of end members that are active in at least one pixel, which coincides with the number of non-zero rows of the abundance matrix $\mathbf{X}$. 
Let $\mathbf{x}^k$ be the $k$th row of matrix $\mathbf{X}$. Then the optimisation problem called \textit{constrained collaborative sparse regression} is solved:
\begin{equation} \label{eq:CLSUnSAL_eq}
    \min_{\mathbf{X}} \|\mathbf{Y}-\mathbf{MX}\|_F^2 + \lambda \sum_{k=1}^{p}\|\mathbf{x}^k\|_2 \quad \text{s.t.} \quad x_{kj} \geq 0 \quad \forall k,j.
\end{equation}
\noindent this method again employs \textit{alternating direction method of multipliers} method and, differently from SUnSAL, it cannot perform pixelwise optimisation, due to the second summand in \eqref{eq:CLSUnSAL_eq}. Moreover, it employs only one parameter, thus easing the parameter tuning.

\noindent A successful improvement of the latter is the so-called \textit{MUSIC-CSR} method, that suitably prunes the spectral library before performing collaborative sparse unmixing \citet{iordache2013music}. As a result, the library mutual coherence is decreased and the unmixing process is speeded up.

\begin{figure}[t]%% placement specifier
%% Use \includegraphics command to insert graphic files. Place graphics files in 
%% working directory.
\centering%% For centre alignment of image.
\includegraphics[width=85mm]{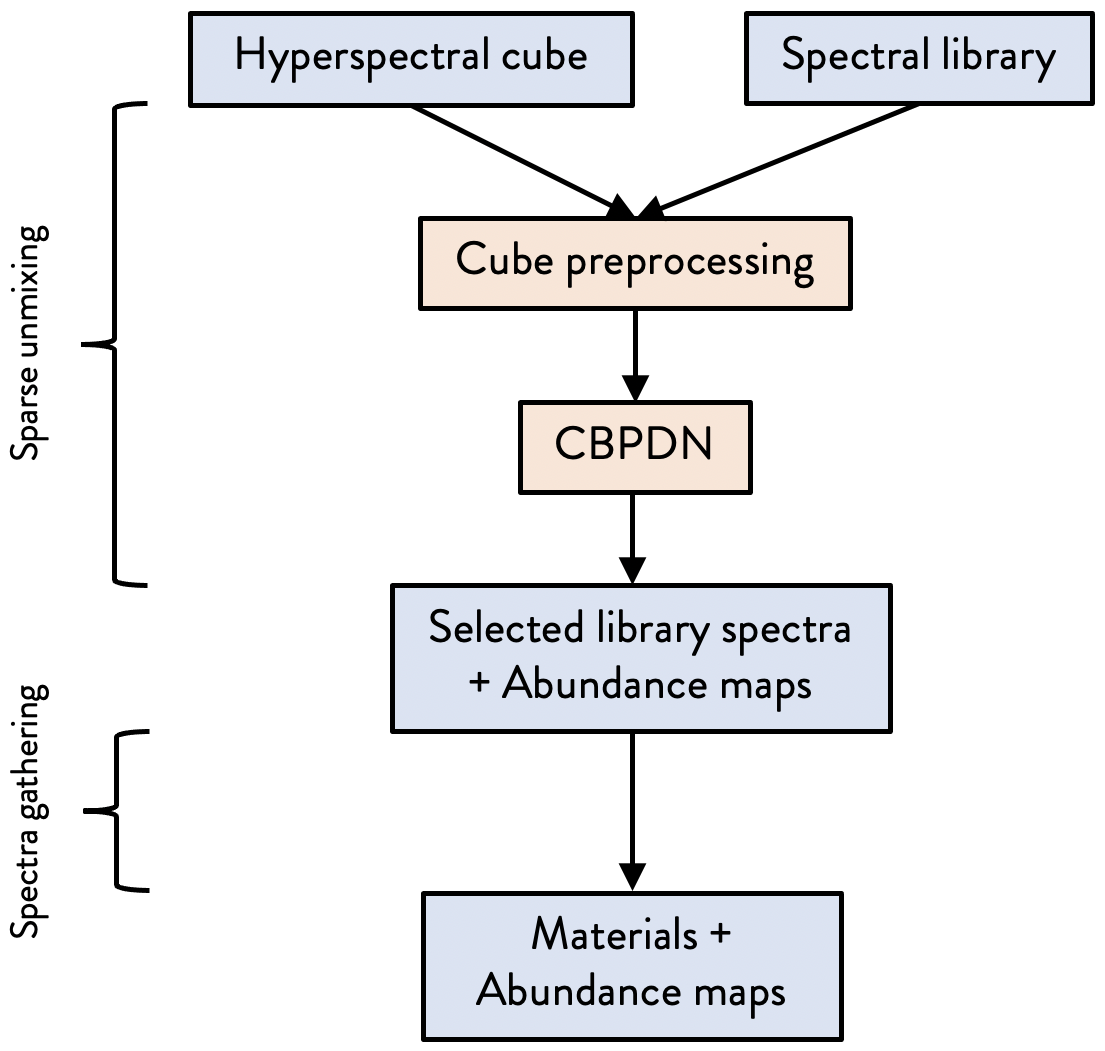}
%% Use \caption command for figure caption and label.
\caption{Workflow for solving the HU problem: sparse unmixing}\label{fig2}
%% https://en.wikibooks.org/wiki/LaTeX/Importing_Graphics#Importing_external_graphics
\end{figure}

In Figure \ref{fig2}, the methodology with sparse unmixing is summarised. Differently from the classical one (cfr. Figure \ref{fig1}, the end members are known, yet the pre-processing step is required, as well as the constrained basis pursuit denoising. In this setting, the mixing models reviewed in Section \ref{sec:classical} are not applicable anymore, so the spectral gathering phase is carried out differently, as explained next.

\subsection{Nonlinear mixing models}
\label{subsec:nonlinear}
\textit{Nonlinear unmixing methods}, mentioned in Section \ref{subsect:Mix_models}, are not as popular as the linear ones \citet{bioucas2012hyperspectral}, but they can prove effective in case LMM does not hold. They can be divided in \textit{data-driven techniques}, that work directly on the data cloud, and \textit{physics-based techniques}, that model the interaction of the light with the ground \citet{heylen2014review}. Data-driven unmixing methods include \textit{neural networks} (NNs), \textit{kernel methods} and \textit{support vector machines} (SVMs), while physics-based methods usually retrieve the abundances under specific mixing models, notably bilinear or intimate ones. 
Focusing on physics-based methods, they usually require the end member signatures to be known a priori. Moreover, they are usually designed just for one mixing model, either bilinear or intimate, thus some prior knowledge of the region to be unmixed is required \citet{bioucas2012hyperspectral}. end member extraction in the nonlinear framework is rarely performed, for instance in \citet{heylen2010non} that exploits \textit{manifold learning}. \\
%In this section, a selection of meaningful nonlinear unmixing methods is provided.

\subsubsection{Intimate mixtures and the Hapke model}

The reader may recall intimate mixtures, introduced in Section \ref{subsect:Mix_models}, which happen when light interacts with particles of different materials several times \citet{heylen2014review}. Such interaction depends on the SSAs (single scattering albedos) of the particles, cfr. Section \ref{subsect:spectroscopy}. 
Though many models describe the light behaviour on particulate surfaces, like \textit{Shkuratov model} \citet{shkuratov1999model}, the most relevant is Hapke model \citet{hapke1981bidirectional}, which provides a relation between \textit{bidirectional reflectance} $\mathbf{y}$ and SSA $\mathbf{w}$. In particular, under further assumptions like the \textit{isotropy} of particles scattering and the particles' sphericity, such relation is well approximated by
\begin{equation} \label{eq:reflectance-SSA}
    y_i = \frac{w_i}{(1+2\mu\sqrt{1-w_i})(1+2\mu_0\sqrt{1-w_i})}
\end{equation}
\noindent where $y_i$, $w_i$ are the components of $\mathbf{y}$ and $\mathbf{w}$ respectively, and $\mu$, $\mu_0$ are the cosines of the angles that the outgoing and incoming radiations form with the normal to the surface respectively. Notably, relation \eqref{eq:reflectance-SSA} is invertible. 
As the SSAs depend only on the first particle hit by light, the SSAs $\{\mathbf{w}_k\}_{k=1}^p$ of materials mix linearly in intimate mixtures \citet{heylen2014review}. Namely, the measured SSA $\mathbf{w}$ at a given pixel is
\begin{equation*}
    \mathbf{w} = \sum_{k=1}^p x_k \mathbf{w}_k
\end{equation*}
\noindent where $x_k$ is the fractional abundance of material $k$ at that pixel. Consequently, a strategy to unmix intimate mixtures is to convert reflectances into SSAs and then perform abundance estimation using the LMM in the SSAs.

\subsubsection{Kernelised linear methods}

Some nonlinear abundance estimation methods employ \textit{kernel functions} to generalise LMM. One notable example is the \textit{kernel fully constrained least squares} (KFCLS) method \citet{broadwater2009comparison, broadwater2007kernel, broadwater2010generalized, broadwater2011mapping}, in which the \textit{scalar products} in the LS objective function \eqref{eq:LMM_LS} are replaced with suitable kernel functions $K(\cdot,\cdot)$, after applying bi-linearity. Consequently, with the above-introduced notation and
%referring to the notation at Subsection \ref{subsect:LMM_eq}, 
dropping the pixel index $j$ for simplicity, the obtained optimisation problem is
\begin{equation} \label{eq:kernel_obj}
        \hat{\mathbf{x}} = \text{arg}\min_{\mathbf{x}} \frac{1}{2} (K(\mathbf{y},\mathbf{y}) - 2\sum_{k=1}^p x_k K(\mathbf{y},\mathbf{m}_k) + \sum_{k=1}^p \sum_{h=1}^p x_k x_h K(\mathbf{m}_k,\mathbf{m}_h)).
\end{equation}
\noindent Abundance non-negativity is usually enforced, while some modifications on \eqref{eq:kernel_obj} are required for the sum constraints \eqref{eq:ASC}. Apart from the classical scalar product and the radial basis function, two interesting kernels are proposed, each including some physical knowledge in the unmixing process. The first is $K^{(1)}(\mathbf{y}_1, \mathbf{y}_2) = \Phi(\mathbf{y}_1)^T \Phi(\mathbf{y}_2)$, a scalar product after a mapping $\Phi(\cdot)$ into the SSA spectrum. $K^{(1)}$ is suitable for unmixing intimate mixtures, as SSAs mix linearly in this scenario. The second kernel is 
\begin{equation}
    K^{(2)}_{\gamma}(\mathbf{y}_1, \mathbf{y}_2) = (1-e^{-\gamma \mathbf{y}_1})^T (1-e^{-\gamma \mathbf{y}_2})
\end{equation}
\noindent where the exponential and the sum are performed component-wise. $K^{(2)}_{\gamma}$ is suitable for linear unmixing for $\gamma$ small, while it is similar to $K^{(1)}$ for $\gamma$ large. $\gamma$ is chosen for each pixel minimising the objective function in \eqref{eq:kernel_obj} jointly with respect to $\gamma$ and $\mathbf{x}$. Thus, $K^{(2)}_{\gamma}$ allows to unmix both linear and intimate mixtures, and the value of $\gamma$ at each pixel suggests the kind of mixing at that pixel. 
A major drawback of this approach is that, despite applying a nonlinear distortion to each end member, it does not keep into account the possible nonlinear interactions between them \citet{chen2012nonlinear}.

\subsection{Advances in hyper-spectral unmixing}\label{subsect:advances}

The most recent directions of research in the hyper-spectral unmixing field are (i) end member variability, and (ii) unmixing with neural networks. These two lines are here overviewed.

%\subsubsection{Spectral variability} %\label{subsect:endm_variability}

{Spectral variability} is the effect for which different measured spectra of the same material may differ \citet{hong2018augmented}. This effect, which may impact negatively on the result of LMM, is usually due to differences in illumination, atmospheric effects and material variability. A complete review article on spectral variability is available \citet{borsoi2021spectral}. 
An unmixing method keeping spectral variability into account is \textit{extended linear mixing model}, that introduces a scaling factor for each end member spectrum at each pixel \citet{drumetz2016blind}.
%\subsubsection{ALMM algorithm} %\label{subsect:ALMM_algo}
The most popular abundance estimation method that considers spectral variability is \textit{augmented linear mixing model} (\citet{hong2018augmented}. It models different sources of spectral variability, including scaling factors. The additional sources are modelled through a set of $\ell$ variability spectra, called \textit{spectral variability dictionary}. 
%Referring to the notation at Subsection \ref{subsect:LMM_eq},
The model is:
\begin{equation}
    \mathbf{Y} = \mathbf{MXS} + \mathbf{EB} + \mathbf{W}
\end{equation}
\noindent where $\mathbf{S}$ is a diagonal matrix containing the scaling factors of each pixel, $\mathbf{E}=[\mathbf{e}_1,...,\mathbf{e}_{\ell}]$ is the spectral variability dictionary and $\mathbf{B} = [\mathbf{b}_1,...,\mathbf{b}_n]$ contains the coefficients of the variability spectra in each pixel. The mixing matrix $\mathbf{M}$ is supposed known, while the remaining matrices are retrieved minimising
\begin{equation} \label{eq:ALMM_minimization}
    \min_{\mathbf{X,B,S,E}} \frac{1}{2}\|\mathbf{Y} - \mathbf{MXS} - \mathbf{EB} \|_{F}^2 + \Phi(\mathbf{X}) + \Psi(\mathbf{B}) + \Gamma(\mathbf{E}) \quad \text{s.t.} \quad 
    \mathbf{X} \geq 0, \mathbf{S} \geq 0
\end{equation}
\noindent where the inequalities are meant component-wise. $\Phi$, $\Psi$, $\Gamma$ summands enforce prior knowledge in the unknowns' estimation. In particular, $\Phi$ promotes sparsity in the abundances and $\Gamma$ pushes $\{ \mathbf{e}_j \}_{j=1}^{\ell}$ to be distinguishable from the employed end member spectra and orthogonal.

%\subsubsection{Neural networks}

Another recent direction of research is the use of Neural Networks (NN) in HU. Two architectures have been mostly inspected up to now. 
The most relevant one is the {autoencoder}, a NN that allows to learn a compressed representation of the input in unsupervised way. It is composed by an {encoder}, that yields such representation, followed by a {decoder}, that reconstructs the original input. These architectures are usually employed both for end member extraction and abundance estimation. 
For instance, \citet{palsson2018hyperspectral} employs just one layer with linear {activation function} for the decoder part, so that the abundances are identified with the compressed representation, while the end member spectra are identified with the weights of the decoder, according to the LMM. A successful instance of autoencoder-based architecture is DAEN, that estimates end members and abundances respecting the abundance non-negativity and sum constraints \citet{su2019daen}. 
A less employed architecture is the {convolutional neural network} (CNN). An example is \citet{zhang2018hyperspectral}, that performs abundance estimation. First, the features are extracted through a CNN, then they are fed to a {multilayer perceptron} and, finally, the outputs of the last layer are scaled to obtain fractional abundances.

More recent applications of different NN architectures which have demonstrated high performance in the context of remote sensing may be promising for the HU problem. They can be utilised to increase the quality of HSIs: for e.g., cascading transformers in \citet{cinese_casformer} have been employed for such end. An anomaly detection point of view could be taken as well. Statistical models, NNs or both could be applied. Indeed, learning disentangled priors have been proposed in \citep{cinese_disentangledpriors}, incorporating both model and data-driven prior knowledge into a neural network for anomaly detection. Another possible direction is that of image segmentation, see \citet{cinese_urban} where adapters are utilised in a deep NN. Whereas the literature of NN architectures in the larger context of remote sensing is being developed with innovative proposals, a critical research direction would be a methodology that incorporates them into the HU problem. One could work with a library of reference spectra as in Subsection \ref{subsect:sparse_unmixing_problem}, and from observed spectra, utilise a classifier to identify the minerals that are present at each pixel. A recent review on deep learning methods for hyper-spectral classification is available \citet{cinese_paraculo_review}, and even recently-developing Mamba architectures have been applied, cfr. \citet{cinese_paraculo_mamba}.

\subsection{Open problems in Hyper-spectral unmixing}\label{subsec:open_probs}

It can be evinced from what has been discussed so far, both in classical and alternative methodologies, that the key components in models which address the HU problem are the following: 
\begin{itemize}
    \item Assumptions on how the minerals are mixed
    \item Choices on the number, type and minerals themselves; whether they are given or estimated from data.
\end{itemize}

In our view, difficult and open problems that are to solve typically lie in one of these two components. These include:
\begin{itemize}
    \item \textbf{Model diagnostics and testability of model assumptions.} As it has been shown throughout this article, models are usually applied by checking whether their suppositions are reasonable in a given context. However, in the literature of Hyper-spectral unmixing, a gap to fill would be to develop either tests that from the available data  quantify the probability of fulfilling them. To our knowledge, only in the case of statistical models it is indirectly satisfied through hypothesis testing, cfr. \citet{cressie2015statistics}.
    \item \textbf{Uncertainty quantification} Whereas also present implicitly in Bayesian models, such element is crucial after having fit a model. In \citet{iordache2011sparse}, for instance, the goodness of the model is demonstrated through simulations and a qualitative check with a known hypercube. Yet when satellites are sent to space, celestial bodies are observed for the first time, and models will be applied insofar as scientists will be able to trust them. 
    \item \textbf{Bridging remote sensing and Hyper-spectral unmixing}. As seen in Subsection \ref{subsect:advances}, a plethora of algorithms and methods are rapidly growing in the closedly related field of remote sensing. The applications are usually different and the objective is not necessarily to unmix minerals, yet they always work with a hyper-spectral cube. We deem tapping into the more general remote sensing field as a challenge that is to be completed.
    \item \textbf{Transfer learning} Neural networks are nowadays ubiquitous in science, yet they require large amounts of data. What about the HU problem, where only a (hyper-spectral) image of a single celestial body is observed? Spatial statistics has usually dealt with such case \citet{cressie2015statistics}, yet a clear methodology of training models putting together many images of different planets appears to us as an important challenge which could be fruitful.
\end{itemize}

% NNs only one image?
% remote sensing applicatio
% uncertainty quantification
% model diagnostics.

%%
%%%%%%%%%%%%%%%%
%%

\section{Datasets review}
\label{sec:datasets}
% TODO 

\subsection{Spectral libraries}

As mentioned several times in the previous Section \ref{sec:Methods_review}, sometimes end members are not to be estimated as reference spectra are provided. A set of available reference spectra, usually measured under similar conditions, is called a spectral library.
Many scientific institutions provide {spectral libraries}. Many libraries include spectra of minerals and rocks, and some are entirely devoted to them. These are particularly suitable for mineralogical applications, which are the main focus of this work. Currently, the most relevant spectral libraries are United States Geological Survey (USGS) and ASTER/ECOSTRESS \citet{xie2022open}. Table \ref{table:Spectral_libraries} gathers the main features of the libraries reviewed in this work.

\begin{table}[H]
    \scriptsize
    \centering 
    \begin{tabular}{|c|l|l|l|l|}
        \hline
        %\rowcolor{bluepoli}
            \textbf{N°} & \textbf{Name} & \textbf{N° of samples} & \textbf{Spectral bands (wavelengths)} \\
        \hline \hline
        01 & USGS Spectral Library Version 7 & $>2400$ (samples) & 0.2-200 $\mu m$ (ultraviolet - far infrared)  \\
        02 & ECOSTRESS (ASTER) spectral library & $>3400$ (samples) & 0.35-15.4 $\mu m$  \\
        03 & ASU spectral library & $>150$ (samples) & 5-45 $\mu m$ (thermal infrared) \\
        04 & Berlin emissivity database (BED) & $>20$ (minerals) & 7-22 $\mu m$ (thermal infrared) \\
        05 & PTAL project & $>100$ (samples) & 0.99-3.6 $\mu m$ (near-infrared) \\
        06 & NMC spectral library & $>590$ (samples) & VIS-NIR-SWIR  \\
        07 & PDS Geosciences Node spectral library & $>11000$ (samples) & UV-Visible-NIR-MIR-FIR \\
        \hline
    \end{tabular}
    \\[10pt]
    \caption{Main features of the most relevant spectral libraries}
    \label{table:Spectral_libraries}
\end{table}

%\subsubsection{USGS Spectral Library Version 7} \label{subseFin pact:USGS}

\begin{figure}[t]%% placement specifier
%% Use \includegraphics command to insert graphic files. Place graphics files in 
%% working directory.
\centering%% For centre alignment of image.
\includegraphics[width=80mm]{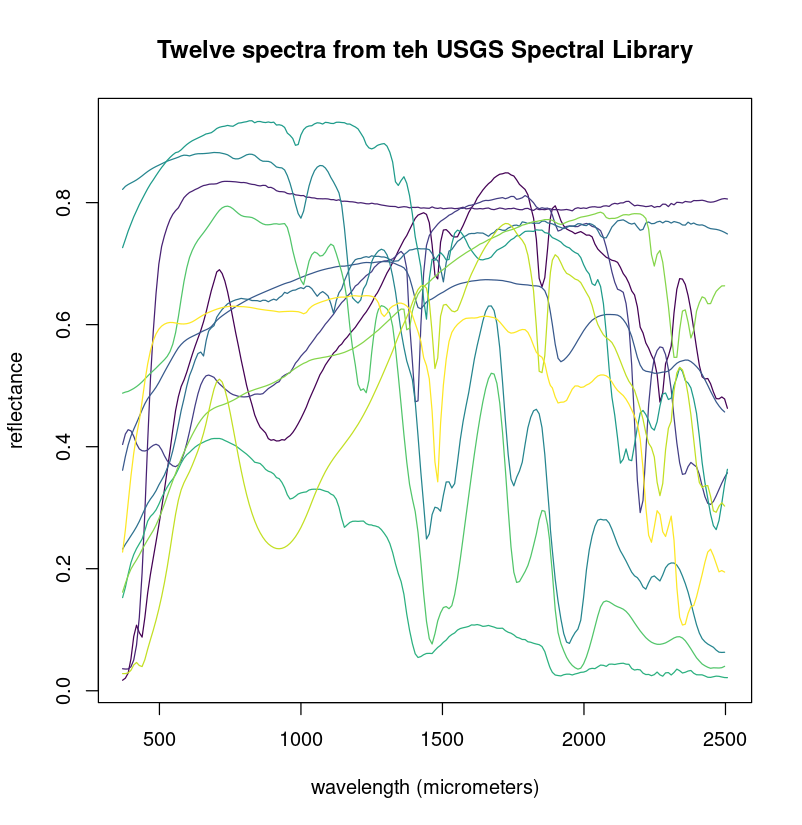}
%% Use \caption command for figure caption and label.
\caption{Plot of 10 randomly chosen spectra from Chapter M of the USGS spectral library}\label{fig3}
%% https://en.wikibooks.org/wiki/LaTeX/Importing_Graphics#Importing_external_graphics
\end{figure}

%\footnote{\url{https://pubs.er.usgs.gov/publication/ds1035}} Version 7} TODO quote properl
USGS Spectral Library 
supports the {United States Geological Survey} (USGS) in its research \citet{kokaly2017usgs}. It is used to interpret spectra from laboratory, field, airborne, orbital instruments, with applications to remote identification and mapping of materials. 
It contains reflectance spectra of many materials, including minerals, rocks, soils, liquids, water, vegetation. Such spectra are measured both through laboratory, field and airborne spectrometers.
Its spectra are adapted to the spectral characteristics of AVIRIS as computed by NASA in 1997, when Cuprite image was acquired (cfr. Section \ref{section:Hyper_cubes}). Consequently, its band wavelengths, provided in $\mu m$, almost coincide with the ones of said image. 
Spectra are grouped into thematic folders called chapters \citet{kokaly2017usgs}, of which only \textit{ChapterM} contains spectra of minerals, mostly in the spectral range from 0.4 $\mu m$ to 2.5 $\mu m$. A sample of spectra from such set are plotted in Figure \ref{fig3}. In some cases, measurements of the same mineral sample with different spectrometers are available. \textit{ChapterS} contains spectra of soils, rocks and mineral mixtures. Moreover, each spectrum has an associated degree of reliability.

%\subsubsection{ECOSTRESS (ASTER) spectral library}
The {ECOSTRESS spectral library} %\footnote{\url{https://speclib.jpl.nasa.gov/}} TODo
O comes from the contribution of three institutions: Johns Hopkins University, \textit{Jet Propulsion Laboratory} (JPL) and USGS. Originally called ASTER library \citet{baldridge2009aster}, it has become the ECOSTRESS library \citet{meerdink2019ecostress} with the inclusion of spectra of vegetation. 
It includes bidirectional and hemispherical reflectance spectra of a wide range of materials, including minerals, rocks, meteorites, lunar and terrestrial soils. Also water, snow, ice and vegetation are included. The library supports ASTER and ECOSTRESS researches, in the field of remote sensing.

%\subsubsection{ASU spectral library}

The \textit{ASU spectral library}
\citet{christensen2000thermal} is provided by the Christensen Research Group at Arizona State University. It contains emission spectra of many minerals, rocks, soils, regoliths and meteorites. 
Such spectra are meant to be compared with spectra measured through planetary spacecrafts or airborne instruments, with a focus on the ones of Martian surface materials, measured by the Thermal Emission Spectrometer (TES) on Mars.

%\subsubsection{Berlin emissivity database}

Antother key library is the {Berlin emissivity database} (BED) \citet{maturilli2008berlin}, which contains emission spectra of many minerals, including many Martian analogues, and also volcanic and lunar highland soils. Its strength is that spectra of each sample are measured for four different particle sizes: $<$25, 25-63, 63-125 and 125-250 $\mu m$. This makes BED more suitable for planetary exploration than ASU, which instead features too large grain sizes (700-1100 $\mu m$).

%\subsubsection{PTAL project}

The {Planetary Terrestrial Analogues Library} %\footnote{\url{https://www.ptal.eu/database-training}} 
(PTAL) \citet{loizeau2020planetary} contains many spectra of Earth rock samples, comprising Martian analogues. It can be applied for remote sensing, with a focus on Mars surface, as the measured samples were chosen while considering the current geological knowledge of Mars.

Another relevant example is the {NMC spectral library} \citet{percival2018customized} includes samples from the {National Mineral Reference Collection} (NMC) of the Geological Survey of Canada. It contains reflectance spectra of many minerals, with a particular focus on pure clay minerals. Many different spectral samples of the same mineral are included, thus providing an idea of the variability of the spectra of a given mineral.

%\subsubsection{PDS Geosciences Node spectral library}
Furthermore, NASA's
\textit{PDS spectral library} %\footnote{\url{https://pds-geosciences.wustl.edu/spectrallibrary/default.htm}} TODO
\citet{pds-dataset}
contains spectra from different institutions, the most relevant of which is the {Reflectance Experiment Laboratory}. It stores spectra of many materials, notably lunar and Mars meteorites, returned lunar samples and terrestrial samples.

\subsection{Hyper-spectral Cubes} \label{section:Hyper_cubes}

Hyper-spectral cubes are the main dataset in the HU mixing problem. As stated above (cfr. Section \ref{sec:Methods_review}), they are an image whose pixels contain (ideally) measured reflectance at different wavelengths.
In most papers on HU, the proposed algorithms are tested both on synthetic data and on {real hyper-spectral cubes.} Some of the latter are quite recurrent, in particular concerning the search of minerals and rocks in arid regions. These are usually similar also to the observed ones in celestial bodies, so in practice HU algorithms are designed for both. Naturally, working with Earthly hypercubes has the advantage of being able to check whether the unmixing algorithm works, whereas direct exploration would be required for celestial bodies.

\subsubsection{AVIRIS imagery} \label{subsect:AVIRIS}

The {Airborne Visible/Infrared Imaging Spectrometer} (AVIRIS)%\footnote{\url{https://aviris.jpl.nasa.gov/index.html}} TODO
, developed at the JPL, is an optical sensor that captures calibrated radiance images in 224 spectral bands. Wavelengths ranges span from 400 $nm$ to 2500 $nm$, with a separation between central bands of around 10 $nm$.
Cubes are acquired mostly in the United States, and some of them are provided also in atmospheric-corrected reflectance \citet{zortea2009spatial}. Most HU algorithms are tested on AVIRIS imagery. 

\begin{figure}[t]%% placement specifier
%% Use \includegraphics command to insert graphic files. Place graphics files in 
%% working directory.
\centering%% For centre alignment of image.
\includegraphics[width=130mm]{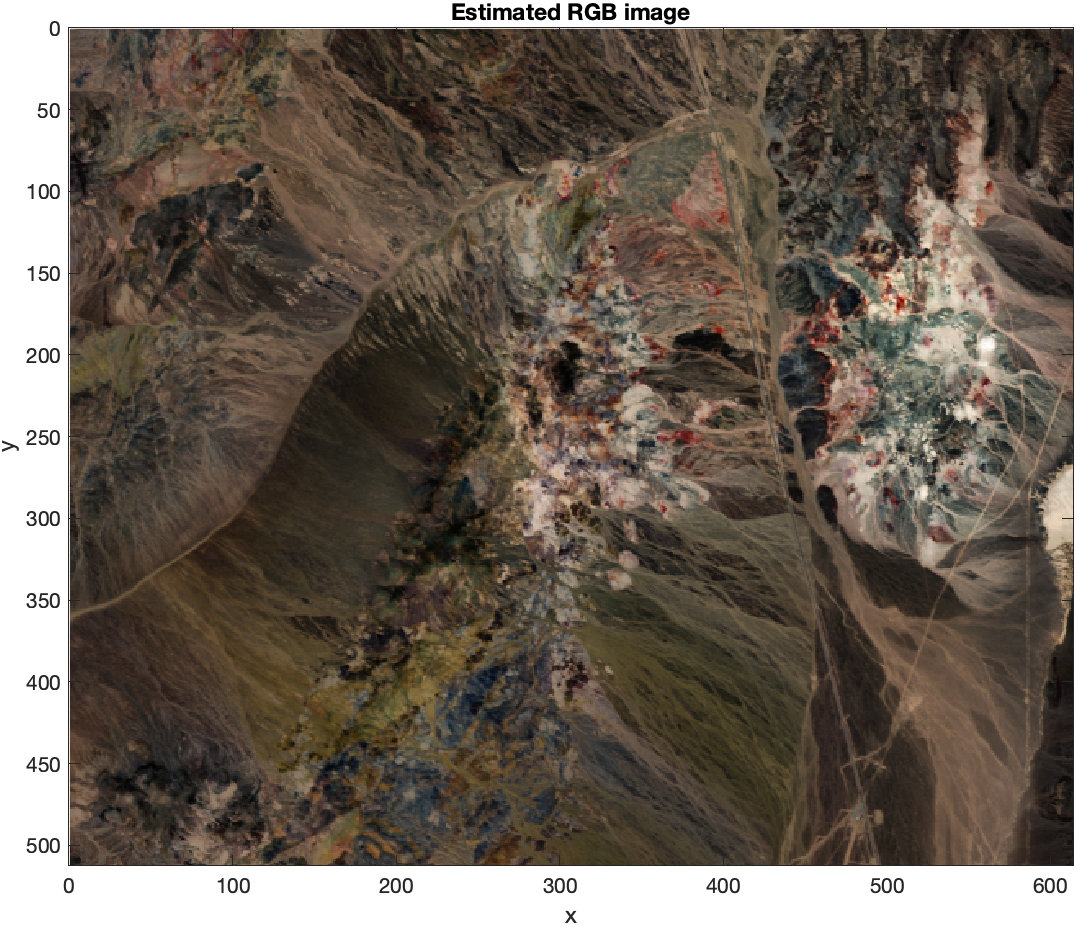}
%% Use \caption command for figure caption and label.
\caption{Visualisation of the AVIRIS Cuprite hyper cube downs caled to three (red, green, blue) colours. Down scaling was implemented with MATLAB Image Processing Toolbox Hyperspectral Imaging Library \citet{matlab}}\label{fig4}
%% https://en.wikibooks.org/wiki/LaTeX/Importing_Graphics#Importing_external_graphics
\end{figure}

In particular, {AVIRIS Cuprite} file %f970619t01p02$\_$r02
cube represents a standard case study in HU, being the most analysed HSI in the  field. It has been acquired in June 1997 on \textit{Cuprite mining area}, Nevada (USA). Each pixel covers an area of approximately 20 $m$ $\times$ $20 m$ \citet{rogge2007integration}.
This HSI is particularly suitable for testing HU methods because Cuprite region features a wide variety of minerals, and is mineralogically well known \citet{swayze1992ground}. The image (using the absorbance at only three wavelengths) is displayed in Figure \ref{fig4}. Moreover, advanced \textit{spectroscopic analyses} of the region employing expert knowledge have been performed by USGS, resulting in \textit{mineral maps} that can be qualitatively compared with the results of this work \citet{swayze2014mapping,clark2003imaging}. 

Other AVIRIS data include the following. 
{AVIRIS Moffett Field} \citet{dobigeon2009joint} covers Moffett Field, California (USA), featuring both rocky and urban areas. Captured in 1997, it is employed for unmixing with both linear and nonlinear methods. {AVIRIS Indian Pines} \citet{nascimento2005does} covers Indian Pine Test Site, Indiana (USA), consisting in a mixture of agricultural and forest ground. It has been captured in 1992, and it has a ground pixel resolution of 17 $m$. It has been applied in few papers dealing with classical unmixing methods. {AVIRIS Lunar Crater Volcanic Field} \citet{heinz2001fully} covers an area in Northern Nye County, Nevada (USA). 
Lastly, \textit{AVIRIS Oatman} \citet{berman2004ice} covers the semiarid Oatman area in Arizona (USA). It features many exposed minerals, that are present in different proportions, and different brightness conditions.

\subsubsection{Mars imagery}

Many HSIs of Mars surface are available, captured by the {Compact Reconnaissance Imaging Spectrometer for Mars} (CRISM) and by the spectrometers of the Observatoire pour la Minéralogie, l’Eau, les Glaces et l’Activité (OMEGA). 

CRISM is meant to map the entire Mars surface \citet{murchie2007compact}. It is on board of the Mars Reconnaissance Orbiter (MRO) spacecraft, launched by NASA in 2005. Images acquired in targeted mode feature a spatial resolution of up to 18 $m$ per pixel and a spectral resolution of 6.5 $nm$ per channel, with channels ranging from 362 to 3920 $nm$ \citet{ceamanos2011intercomparison}. Moreover, a spectral library of Martian analogues is available online % \footnote{\url{https://pds-geosciences.wustl.edu/missions/mro/spectral_library.htm}} TODO
for comparison with CRISM imagery. 

A typically used CRISM image is the one named \textit{frt000042aa}, acquired in July 2009. It covers the Mars Russell mega dune, which has been carefully studied. Moreover, high spatial resolution images of the region have been acquired through HiRISE camera, and they have been used to build a reliable ground truth to test the unmixing methods \citet{ceamanos2011intercomparison}. 

OMEGA is on the Mars Express (MEX) orbiter of ESA, launched in 2003 \citet{bibring2004omega,liu2018exploration}. It features 352 contiguous bands in the range 0.38-5.1 $\mu m$ and a spatial resolution of less than 350 $m$ per pixel. \\
An instance of OMEGA image is \textbf{ORB0041} image, covering Mars south polar cap. This area has been carefully studied, concluding that 3 main chemical species are present ($\text{H}_2\text{O}$ ice, $\text{CO}_2$ ice and mineral dust) and that intimate mixtures dominate the area.

\subsubsection{Further datasets}

There are other well-renowned datasets which despite being used also in some unmixing papers, are probably more suitable for pixel-wise classification techniques, mostly due to their high spatial resolution. Moreover, they cover urban areas, so they are not suitable case studies for mineral research. 
Among these, we mention
\textit{Urban} \citet{zhu2014structured} that was acquired in October 1995 by HYDICE and covers the area of Copperas Cove, Texas (USA). It is oftentimes employed in HU papers, also in most recent ones dealing with NNs and NMF, despite its high spatial resolution (2 $m$ per pixel). Finally, \textit{CASI Houston}, acquired in June 2012 by CASI imager, covers the University of Houston campus with 2.5 $m$ of spatial resolution, while \textit{ROSIS Pavia}, acquired by ROSIS-03 airborne instrument, covers the University of Pavia campus \citet{ghamisi2017advances}. Both are rarely employed in unmixing papers.

\section{Discussion and further outlooks}\label{sec:concl}

In this work we have provided an updated literature review about hyper-spectral unmixing. We started with the problem definition, i.e. stating what a mixing model is and characterising it by its constraints. Next, we reviewed the classical workflow: linear mixing is assumed, and using end members --which in turn can be estimated in several ways-- fractional abundances at each pixel are estimated. We also covered alternative methodologies, namely sparse unmixing, nonlinear mixing and the utilisation of other techniques such as statistical models and neural networks. Such section was followed by a survey of the most important datasets in this setting, which are the hyper-spectral cubes themselves and the spectral libraries, which are not strictly necessary but provide a reference of the end members to look for, instead of extracting them from the data. 

\subsection{Summary}
We deem relevant to share the following remarks:
\begin{itemize}
    \item Hyper-spectral unmixing is a problem which is pertinent to different researchers and is in that sinterdisciplinaryiplinary problem. Physical, statistical, information-theoretic, geological  models, among others, are employed. Efficient and effective algorithms have been devised and others are being proposed for their computation. Therefore, a correct approach to solve this problem must be a global one which considers contributions from all the fields that address it.
    \item A key assumption in the classical methodology (see Section \ref{sec:Methods_review}) is the linear-mixing, which is guaranteed, e.g., in  the case of areal mixtures. Whereas it is convenient for its simplicity and interpretability, in the literature it has been verified that such model does not necessarily reflect the actual phenomenon. Hence, a step further in the direction of nonlinear mixing should be encouraged, although we recognise the challenge since most of the literature focuses on linear mixing.
    \item We noticed that a significant number of end member extraction algorithms work with the pure-pixel hypothesis. We saw in Section \ref{subsec:end member-no-pixel} procedures which do not have such assumption, and in particular \citet{plaza2002spatial} use classic image processing for both end member extraction and pure pixel evaluation. Methods that explicitly evaluate the pure-pixel assumption should be explored further. This may also include, e.g., specific statistical testing on the pure-pixel hypothesis.
    % TODO qualcosa: neural networks use lots of images; spatial statistics one realisation...
\end{itemize}

\subsection{Recommendations for future research}

The open problems we have identified were (cfr. Subsection \ref{subsec:open_probs}): {model diagnostics and testability of model assumptions}, {uncertainty quantification}, {bridging further the HU literature with the broader remote sensing literature}, and transfer learning.
\begin{itemize}
    \item The uncertainty in the outputs of the models and/or algorithms applied to hyper-spectral cubes is seldom shown. Yet, this is valuable information to evaluate the goodness of a model in the absence of a ground truth, which is generally missing, apart from the Cuprite and other very specific cases. While (parametric) statistical models such as linear mixing fitted with (fully) constrained least squares tend to provide estimates of the variance of the parameters' estimates, the literature has been poor in terms of uncertainty quantification. We see this as a broad yet almost unexplored research field. A solution could be adapting \citet{gelfand}'s work, who work with a similar data set. At each pixel, they observe a reflectance curve. Through Bayesian models for both these functions and their spatial dependence, they provide an estimate for their distribution, as well as the uncertainty quantification in the fitted parameters. {Simulating from the fitted model could yield data to combine with an endmember selection method, having as a by-product the uncertainty estimate of the Monte Carlo method}.
    
    \item As a last remark, and knowing that hyper-spectral unmixing is indeed an interdisciplinary problem, the intersection of spatial statistics, image processing and computer vision should be further investigated. For example, neural networks have been used for this problem (see Section \ref{subsec:nonlinear}). Combining NNs and statistics has been recently done in \citet{cressiespatialbayesianneuralnetworks}, who obtain uncertainty estimates besides (abundance) predictions. Ideally, {a library of Hyper-spectral cubes could be created, and deep (Bayesian) Neural Networks could be trained on them}. Then, {the pipeline upon receiving a new hypercube of a new celestial body, for example, would just require fine-tuning the learnt network to the latest cube}. As such, the integration of these fields, in the view of the authors, have clear potential for an improved modelling and estimation framework in the context of this work.
\end{itemize}

%
%\section*{CRediT authorship contribution statement}
%  \textbf{Alessandra Menafoglio}:. \textbf{Simone Vantini}: \textbf{Alfredo Gimenez Zapiola}: \textbf{Andrea Boselli}

\section*{Declaration of competing interest}
The authors declare that they have no known competing financial interests or personal relationships that could have appeared to influence the work reported in this paper.

\section*{Acknowledgments}
This work has been partially supported by ASI-POLIMI “Attività di Ricerca e Innovazione” project, n. 2018-5-HH.0, a collaboration agreement between the Italian Space Agency (ASI) and Politecnico di Milano

\section*{Data availability}
Whereas no data were used directly in this article, the mentioned datasets are available by request if not already public at the links we have provided in the references.




%\bibliography{Thesis_bibliography} % The references information are stored in the file named "Thesis_bibliography.bib"

\bibliographystyle{elsarticle-num-names}
\bibliography{Thesis_bibliography.bib}

\begin{thebibliography}{82}
\expandafter\ifx\csname natexlab\endcsname\relax\def\natexlab#1{#1}\fi
\providecommand{\url}[1]{\texttt{#1}}
\providecommand{\href}[2]{#2}
\providecommand{\path}[1]{#1}
\providecommand{\DOIprefix}{doi:}
\providecommand{\ArXivprefix}{arXiv:}
\providecommand{\URLprefix}{URL: }
\providecommand{\Pubmedprefix}{pmid:}
\providecommand{\doi}[1]{\href{http://dx.doi.org/#1}{\path{#1}}}
\providecommand{\Pubmed}[1]{\href{pmid:#1}{\path{#1}}}
\providecommand{\bibinfo}[2]{#2}
\ifx\xfnm\relax \def\xfnm[#1]{\unskip,\space#1}\fi
%Type = Article
\bibitem[{Shaw and Burke(2003)}]{shaw2003spectral}
\bibinfo{author}{G.~A. Shaw}, \bibinfo{author}{H.~K. Burke},
\newblock \bibinfo{title}{Spectral imaging for remote sensing},
\newblock \bibinfo{journal}{Lincoln laboratory journal} \bibinfo{volume}{14} (\bibinfo{year}{2003}) \bibinfo{pages}{3--28}.
%Type = Article
\bibitem[{Bioucas-Dias et~al.(2012)Bioucas-Dias, Plaza, Dobigeon, Parente, Du, Gader, and Chanussot}]{bioucas2012hyperspectral}
\bibinfo{author}{J.~M. Bioucas-Dias}, \bibinfo{author}{A.~Plaza}, \bibinfo{author}{N.~Dobigeon}, \bibinfo{author}{M.~Parente}, \bibinfo{author}{Q.~Du}, \bibinfo{author}{P.~Gader}, \bibinfo{author}{J.~Chanussot},
\newblock \bibinfo{title}{Hyperspectral unmixing overview: Geometrical, statistical, and sparse regression-based approaches},
\newblock \bibinfo{journal}{IEEE journal of selected topics in applied earth observations and remote sensing} \bibinfo{volume}{5} (\bibinfo{year}{2012}) \bibinfo{pages}{354--379}.
%Type = Inproceedings
\bibitem[{Broadwater and Banerjee(2010)}]{broadwater2010generalized}
\bibinfo{author}{J.~Broadwater}, \bibinfo{author}{A.~Banerjee},
\newblock \bibinfo{title}{A generalized kernel for areal and intimate mixtures},
\newblock in: \bibinfo{booktitle}{2010 2nd Workshop on Hyperspectral Image and Signal Processing: Evolution in Remote Sensing}, \bibinfo{organization}{IEEE}, \bibinfo{year}{2010}, pp. \bibinfo{pages}{1--4}.
%Type = Article
\bibitem[{Heylen et~al.(2014)Heylen, Parente, and Gader}]{heylen2014review}
\bibinfo{author}{R.~Heylen}, \bibinfo{author}{M.~Parente}, \bibinfo{author}{P.~Gader},
\newblock \bibinfo{title}{A review of nonlinear hyperspectral unmixing methods},
\newblock \bibinfo{journal}{IEEE Journal of Selected Topics in Applied Earth Observations and Remote Sensing} \bibinfo{volume}{7} (\bibinfo{year}{2014}) \bibinfo{pages}{1844--1868}.
%Type = Article
\bibitem[{Ghamisi et~al.(2017)Ghamisi, Yokoya, Li, Liao, Liu, Plaza, Rasti, and Plaza}]{ghamisi2017advances}
\bibinfo{author}{P.~Ghamisi}, \bibinfo{author}{N.~Yokoya}, \bibinfo{author}{J.~Li}, \bibinfo{author}{W.~Liao}, \bibinfo{author}{S.~Liu}, \bibinfo{author}{J.~Plaza}, \bibinfo{author}{B.~Rasti}, \bibinfo{author}{A.~Plaza},
\newblock \bibinfo{title}{Advances in hyperspectral image and signal processing: A comprehensive overview of the state of the art},
\newblock \bibinfo{journal}{IEEE Geoscience and Remote Sensing Magazine} \bibinfo{volume}{5} (\bibinfo{year}{2017}) \bibinfo{pages}{37--78}.
%Type = Article
\bibitem[{Gao et~al.(1993)Gao, Heidebrecht, and Goetz}]{ATREM}
\bibinfo{author}{B.-C. Gao}, \bibinfo{author}{K.~B. Heidebrecht}, \bibinfo{author}{A.~F. Goetz},
\newblock \bibinfo{title}{Derivation of scaled surface reflectances from aviris data},
\newblock \bibinfo{journal}{Remote Sensing of Environment} \bibinfo{volume}{44} (\bibinfo{year}{1993}) \bibinfo{pages}{165--178}. \URLprefix \url{https://www.sciencedirect.com/science/article/pii/003442579390014O}. \DOIprefix\doi{https://doi.org/10.1016/0034-4257(93)90014-O}, \bibinfo{note}{airbone Imaging Spectrometry}.
%Type = Article
\bibitem[{Iordache et~al.(2011)Iordache, Bioucas-Dias, and Plaza}]{iordache2011sparse}
\bibinfo{author}{M.-D. Iordache}, \bibinfo{author}{J.~M. Bioucas-Dias}, \bibinfo{author}{A.~Plaza},
\newblock \bibinfo{title}{Sparse unmixing of hyperspectral data},
\newblock \bibinfo{journal}{IEEE Transactions on Geoscience and Remote Sensing} \bibinfo{volume}{49} (\bibinfo{year}{2011}) \bibinfo{pages}{2014--2039}.
%Type = Article
\bibitem[{Miao and Qi(2007)}]{miao2007endmember}
\bibinfo{author}{L.~Miao}, \bibinfo{author}{H.~Qi},
\newblock \bibinfo{title}{Endmember extraction from highly mixed data using minimum volume constrained nonnegative matrix factorization},
\newblock \bibinfo{journal}{IEEE Transactions on Geoscience and Remote Sensing} \bibinfo{volume}{45} (\bibinfo{year}{2007}) \bibinfo{pages}{765--777}.
%Type = Article
\bibitem[{Bioucas-Dias et~al.(2013)Bioucas-Dias, Plaza, Camps-Valls, Scheunders, Nasrabadi, and Chanussot}]{bioucas2013hyperspectral}
\bibinfo{author}{J.~M. Bioucas-Dias}, \bibinfo{author}{A.~Plaza}, \bibinfo{author}{G.~Camps-Valls}, \bibinfo{author}{P.~Scheunders}, \bibinfo{author}{N.~Nasrabadi}, \bibinfo{author}{J.~Chanussot},
\newblock \bibinfo{title}{Hyperspectral remote sensing data analysis and future challenges},
\newblock \bibinfo{journal}{IEEE Geoscience and remote sensing magazine} \bibinfo{volume}{1} (\bibinfo{year}{2013}) \bibinfo{pages}{6--36}.
%Type = Article
\bibitem[{Hapke(1981)}]{hapke1981bidirectional}
\bibinfo{author}{B.~Hapke},
\newblock \bibinfo{title}{Bidirectional reflectance spectroscopy: 1. theory},
\newblock \bibinfo{journal}{Journal of Geophysical Research: Solid Earth} \bibinfo{volume}{86} (\bibinfo{year}{1981}) \bibinfo{pages}{3039--3054}.
%Type = Article
\bibitem[{Heylen and Scheunders(2015)}]{heylen2015multilinear}
\bibinfo{author}{R.~Heylen}, \bibinfo{author}{P.~Scheunders},
\newblock \bibinfo{title}{A multilinear mixing model for nonlinear spectral unmixing},
\newblock \bibinfo{journal}{IEEE transactions on geoscience and remote sensing} \bibinfo{volume}{54} (\bibinfo{year}{2015}) \bibinfo{pages}{240--251}.
%Type = Article
\bibitem[{Keshava and Mustard(2002)}]{keshava2002spectral}
\bibinfo{author}{N.~Keshava}, \bibinfo{author}{J.~F. Mustard},
\newblock \bibinfo{title}{Spectral unmixing},
\newblock \bibinfo{journal}{IEEE signal processing magazine} \bibinfo{volume}{19} (\bibinfo{year}{2002}) \bibinfo{pages}{44--57}.
%Type = Article
\bibitem[{Dobigeon et~al.(2013)Dobigeon, Tourneret, Richard, Bermudez, McLaughlin, and Hero}]{dobigeon2013nonlinear}
\bibinfo{author}{N.~Dobigeon}, \bibinfo{author}{J.-Y. Tourneret}, \bibinfo{author}{C.~Richard}, \bibinfo{author}{J.~C.~M. Bermudez}, \bibinfo{author}{S.~McLaughlin}, \bibinfo{author}{A.~O. Hero},
\newblock \bibinfo{title}{Nonlinear unmixing of hyperspectral images: Models and algorithms},
\newblock \bibinfo{journal}{IEEE Signal processing magazine} \bibinfo{volume}{31} (\bibinfo{year}{2013}) \bibinfo{pages}{82--94}.
%Type = Article
\bibitem[{Ambikapathi et~al.(2011)Ambikapathi, Chan, Ma, and Chi}]{ambikapathi2011chance}
\bibinfo{author}{A.~Ambikapathi}, \bibinfo{author}{T.-H. Chan}, \bibinfo{author}{W.-K. Ma}, \bibinfo{author}{C.-Y. Chi},
\newblock \bibinfo{title}{Chance-constrained robust minimum-volume enclosing simplex algorithm for hyperspectral unmixing},
\newblock \bibinfo{journal}{IEEE Transactions on Geoscience and Remote Sensing} \bibinfo{volume}{49} (\bibinfo{year}{2011}) \bibinfo{pages}{4194--4209}.
%Type = Article
\bibitem[{Chan et~al.(2009)Chan, Chi, Huang, and Ma}]{chan2009convex}
\bibinfo{author}{T.-H. Chan}, \bibinfo{author}{C.-Y. Chi}, \bibinfo{author}{Y.-M. Huang}, \bibinfo{author}{W.-K. Ma},
\newblock \bibinfo{title}{A convex analysis-based minimum-volume enclosing simplex algorithm for hyperspectral unmixing},
\newblock \bibinfo{journal}{IEEE Transactions on Signal Processing} \bibinfo{volume}{57} (\bibinfo{year}{2009}) \bibinfo{pages}{4418--4432}.
%Type = Article
\bibitem[{Chan et~al.(2011)Chan, Ma, Ambikapathi, and Chi}]{chan2011simplex}
\bibinfo{author}{T.-H. Chan}, \bibinfo{author}{W.-K. Ma}, \bibinfo{author}{A.~Ambikapathi}, \bibinfo{author}{C.-Y. Chi},
\newblock \bibinfo{title}{A simplex volume maximization framework for hyperspectral endmember extraction},
\newblock \bibinfo{journal}{IEEE Transactions on Geoscience and Remote Sensing} \bibinfo{volume}{49} (\bibinfo{year}{2011}) \bibinfo{pages}{4177--4193}.
%Type = Article
\bibitem[{Halimi et~al.(2011)Halimi, Altmann, Dobigeon, and Tourneret}]{halimi2011nonlinear}
\bibinfo{author}{A.~Halimi}, \bibinfo{author}{Y.~Altmann}, \bibinfo{author}{N.~Dobigeon}, \bibinfo{author}{J.-Y. Tourneret},
\newblock \bibinfo{title}{Nonlinear unmixing of hyperspectral images using a generalized bilinear model},
\newblock \bibinfo{journal}{IEEE Transactions on Geoscience and Remote Sensing} \bibinfo{volume}{49} (\bibinfo{year}{2011}) \bibinfo{pages}{4153--4162}.
%Type = Article
\bibitem[{Tao et~al.(2021)Tao, Paoletti, Haut, Ren, Plaza, and Plaza}]{tao2021endmember}
\bibinfo{author}{X.~Tao}, \bibinfo{author}{M.~E. Paoletti}, \bibinfo{author}{J.~M. Haut}, \bibinfo{author}{P.~Ren}, \bibinfo{author}{J.~Plaza}, \bibinfo{author}{A.~Plaza},
\newblock \bibinfo{title}{Endmember estimation with maximum distance analysis},
\newblock \bibinfo{journal}{Remote Sensing} \bibinfo{volume}{13} (\bibinfo{year}{2021}) \bibinfo{pages}{713}.
%Type = Article
\bibitem[{Chang and Du(2004)}]{chang2004estimation}
\bibinfo{author}{C.-I. Chang}, \bibinfo{author}{Q.~Du},
\newblock \bibinfo{title}{Estimation of number of spectrally distinct signal sources in hyperspectral imagery},
\newblock \bibinfo{journal}{IEEE Transactions on geoscience and remote sensing} \bibinfo{volume}{42} (\bibinfo{year}{2004}) \bibinfo{pages}{608--619}.
%Type = Article
\bibitem[{Ambikapathi et~al.(2012)Ambikapathi, Chan, Chi, and Keizer}]{ambikapathi2012hyperspectral}
\bibinfo{author}{A.~Ambikapathi}, \bibinfo{author}{T.-H. Chan}, \bibinfo{author}{C.-Y. Chi}, \bibinfo{author}{K.~Keizer},
\newblock \bibinfo{title}{Hyperspectral data geometry-based estimation of number of endmembers using p-norm-based pure pixel identification algorithm},
\newblock \bibinfo{journal}{IEEE Transactions on Geoscience and Remote Sensing} \bibinfo{volume}{51} (\bibinfo{year}{2012}) \bibinfo{pages}{2753--2769}.
%Type = Article
\bibitem[{Luo et~al.(2012)Luo, Chanussot, Dout{\'e}, and Zhang}]{luo2012empirical}
\bibinfo{author}{B.~Luo}, \bibinfo{author}{J.~Chanussot}, \bibinfo{author}{S.~Dout{\'e}}, \bibinfo{author}{L.~Zhang},
\newblock \bibinfo{title}{Empirical automatic estimation of the number of endmembers in hyperspectral images},
\newblock \bibinfo{journal}{IEEE Geoscience and Remote Sensing Letters} \bibinfo{volume}{10} (\bibinfo{year}{2012}) \bibinfo{pages}{24--28}.
%Type = Article
\bibitem[{Bioucas-Dias and Nascimento(2008)}]{bioucas2008hyperspectral}
\bibinfo{author}{J.~M. Bioucas-Dias}, \bibinfo{author}{J.~M. Nascimento},
\newblock \bibinfo{title}{Hyperspectral subspace identification},
\newblock \bibinfo{journal}{IEEE Transactions on Geoscience and Remote Sensing} \bibinfo{volume}{46} (\bibinfo{year}{2008}) \bibinfo{pages}{2435--2445}.
%Type = Article
\bibitem[{Nascimento and Dias(2005)}]{nascimento2005vertex}
\bibinfo{author}{J.~M. Nascimento}, \bibinfo{author}{J.~M. Dias},
\newblock \bibinfo{title}{Vertex component analysis: A fast algorithm to unmix hyperspectral data},
\newblock \bibinfo{journal}{IEEE transactions on Geoscience and Remote Sensing} \bibinfo{volume}{43} (\bibinfo{year}{2005}) \bibinfo{pages}{898--910}.
%Type = Inproceedings
\bibitem[{Winter(1999)}]{winter1999n}
\bibinfo{author}{M.~E. Winter},
\newblock \bibinfo{title}{N-findr: An algorithm for fast autonomous spectral end-member determination in hyperspectral data},
\newblock in: \bibinfo{booktitle}{Imaging Spectrometry V}, volume \bibinfo{volume}{3753}, \bibinfo{organization}{SPIE}, \bibinfo{year}{1999}, pp. \bibinfo{pages}{266--275}.
%Type = Article
\bibitem[{Chang et~al.(2006)Chang, Wu, Liu, and Ouyang}]{chang2006new}
\bibinfo{author}{C.-I. Chang}, \bibinfo{author}{C.-C. Wu}, \bibinfo{author}{W.~Liu}, \bibinfo{author}{Y.-C. Ouyang},
\newblock \bibinfo{title}{A new growing method for simplex-based endmember extraction algorithm},
\newblock \bibinfo{journal}{IEEE transactions on geoscience and remote sensing} \bibinfo{volume}{44} (\bibinfo{year}{2006}) \bibinfo{pages}{2804--2819}.
%Type = Article
\bibitem[{Boardman et~al.(1995)Boardman, Kruse, and Green}]{boardman1995mapping}
\bibinfo{author}{J.~W. Boardman}, \bibinfo{author}{F.~A. Kruse}, \bibinfo{author}{R.~O. Green},
\newblock \bibinfo{title}{Mapping target signatures via partial unmixing of aviris data},
\newblock \bibinfo{journal}{NASA Technical Report}  (\bibinfo{year}{1995}).
%Type = Article
\bibitem[{Berman et~al.(2004)Berman, Kiiveri, Lagerstrom, Ernst, Dunne, and Huntington}]{berman2004ice}
\bibinfo{author}{M.~Berman}, \bibinfo{author}{H.~Kiiveri}, \bibinfo{author}{R.~Lagerstrom}, \bibinfo{author}{A.~Ernst}, \bibinfo{author}{R.~Dunne}, \bibinfo{author}{J.~F. Huntington},
\newblock \bibinfo{title}{Ice: A statistical approach to identifying endmembers in hyperspectral images},
\newblock \bibinfo{journal}{IEEE transactions on Geoscience and Remote Sensing} \bibinfo{volume}{42} (\bibinfo{year}{2004}) \bibinfo{pages}{2085--2095}.
%Type = Article
\bibitem[{Harsanyi and Chang(1994)}]{harsanyi1994hyperspectral}
\bibinfo{author}{J.~C. Harsanyi}, \bibinfo{author}{C.-I. Chang},
\newblock \bibinfo{title}{Hyperspectral image classification and dimensionality reduction: An orthogonal subspace projection approach},
\newblock \bibinfo{journal}{IEEE Transactions on geoscience and remote sensing} \bibinfo{volume}{32} (\bibinfo{year}{1994}) \bibinfo{pages}{779--785}.
%Type = Article
\bibitem[{Qian et~al.(2011)Qian, Jia, Zhou, and Robles-Kelly}]{qian2011hyperspectral}
\bibinfo{author}{Y.~Qian}, \bibinfo{author}{S.~Jia}, \bibinfo{author}{J.~Zhou}, \bibinfo{author}{A.~Robles-Kelly},
\newblock \bibinfo{title}{Hyperspectral unmixing via $ l\_ $\{$1/2$\}$ $ sparsity-constrained nonnegative matrix factorization},
\newblock \bibinfo{journal}{IEEE Transactions on Geoscience and Remote Sensing} \bibinfo{volume}{49} (\bibinfo{year}{2011}) \bibinfo{pages}{4282--4297}.
%Type = Article
\bibitem[{Yao et~al.(2019)Yao, Meng, Zhao, Cao, and Xu}]{yao2019nonconvex}
\bibinfo{author}{J.~Yao}, \bibinfo{author}{D.~Meng}, \bibinfo{author}{Q.~Zhao}, \bibinfo{author}{W.~Cao}, \bibinfo{author}{Z.~Xu},
\newblock \bibinfo{title}{Nonconvex-sparsity and nonlocal-smoothness-based blind hyperspectral unmixing},
\newblock \bibinfo{journal}{IEEE Transactions on Image Processing} \bibinfo{volume}{28} (\bibinfo{year}{2019}) \bibinfo{pages}{2991--3006}.
%Type = Inproceedings
\bibitem[{Bioucas-Dias(2009)}]{bioucas2009variable}
\bibinfo{author}{J.~M. Bioucas-Dias},
\newblock \bibinfo{title}{A variable splitting augmented lagrangian approach to linear spectral unmixing},
\newblock in: \bibinfo{booktitle}{2009 First workshop on hyperspectral image and signal processing: Evolution in remote sensing}, \bibinfo{organization}{IEEE}, \bibinfo{year}{2009}, pp. \bibinfo{pages}{1--4}.
%Type = Article
\bibitem[{Li et~al.(2015)Li, Agathos, Zaharie, Bioucas-Dias, Plaza, and Li}]{li2015minimum}
\bibinfo{author}{J.~Li}, \bibinfo{author}{A.~Agathos}, \bibinfo{author}{D.~Zaharie}, \bibinfo{author}{J.~M. Bioucas-Dias}, \bibinfo{author}{A.~Plaza}, \bibinfo{author}{X.~Li},
\newblock \bibinfo{title}{Minimum volume simplex analysis: A fast algorithm for linear hyperspectral unmixing},
\newblock \bibinfo{journal}{IEEE Transactions on Geoscience and Remote Sensing} \bibinfo{volume}{53} (\bibinfo{year}{2015}) \bibinfo{pages}{5067--5082}.
%Type = Article
\bibitem[{Dobigeon et~al.(2009)Dobigeon, Moussaoui, Coulon, Tourneret, and Hero}]{dobigeon2009joint}
\bibinfo{author}{N.~Dobigeon}, \bibinfo{author}{S.~Moussaoui}, \bibinfo{author}{M.~Coulon}, \bibinfo{author}{J.-Y. Tourneret}, \bibinfo{author}{A.~O. Hero},
\newblock \bibinfo{title}{Joint bayesian endmember extraction and linear unmixing for hyperspectral imagery},
\newblock \bibinfo{journal}{IEEE Transactions on Signal Processing} \bibinfo{volume}{57} (\bibinfo{year}{2009}) \bibinfo{pages}{4355--4368}.
%Type = Article
\bibitem[{Zare and Gader(2010)}]{zare2010pce}
\bibinfo{author}{A.~Zare}, \bibinfo{author}{P.~Gader},
\newblock \bibinfo{title}{Pce: Piecewise convex endmember detection},
\newblock \bibinfo{journal}{IEEE Transactions on Geoscience and Remote Sensing} \bibinfo{volume}{48} (\bibinfo{year}{2010}) \bibinfo{pages}{2620--2632}.
%Type = Article
\bibitem[{Xu et~al.(2018)Xu, Li, Wu, and Plaza}]{xu2018regional}
\bibinfo{author}{X.~Xu}, \bibinfo{author}{J.~Li}, \bibinfo{author}{C.~Wu}, \bibinfo{author}{A.~Plaza},
\newblock \bibinfo{title}{Regional clustering-based spatial preprocessing for hyperspectral unmixing},
\newblock \bibinfo{journal}{Remote Sensing of Environment} \bibinfo{volume}{204} (\bibinfo{year}{2018}) \bibinfo{pages}{333--346}.
%Type = Article
\bibitem[{Plaza et~al.(2002)Plaza, Martinez, P{\'e}rez, and Plaza}]{plaza2002spatial}
\bibinfo{author}{A.~Plaza}, \bibinfo{author}{P.~Martinez}, \bibinfo{author}{R.~P{\'e}rez}, \bibinfo{author}{J.~Plaza},
\newblock \bibinfo{title}{Spatial/spectral endmember extraction by multidimensional morphological operations},
\newblock \bibinfo{journal}{IEEE transactions on geoscience and remote sensing} \bibinfo{volume}{40} (\bibinfo{year}{2002}) \bibinfo{pages}{2025--2041}.
%Type = Article
\bibitem[{Rogge et~al.(2007)Rogge, Rivard, Zhang, Sanchez, Harris, and Feng}]{rogge2007integration}
\bibinfo{author}{D.~M. Rogge}, \bibinfo{author}{B.~Rivard}, \bibinfo{author}{J.~Zhang}, \bibinfo{author}{A.~Sanchez}, \bibinfo{author}{J.~Harris}, \bibinfo{author}{J.~Feng},
\newblock \bibinfo{title}{Integration of spatial--spectral information for the improved extraction of endmembers},
\newblock \bibinfo{journal}{Remote Sensing of Environment} \bibinfo{volume}{110} (\bibinfo{year}{2007}) \bibinfo{pages}{287--303}.
%Type = Article
\bibitem[{Zortea and Plaza(2009)}]{zortea2009spatial}
\bibinfo{author}{M.~Zortea}, \bibinfo{author}{A.~Plaza},
\newblock \bibinfo{title}{Spatial preprocessing for endmember extraction},
\newblock \bibinfo{journal}{IEEE Transactions on Geoscience and Remote Sensing} \bibinfo{volume}{47} (\bibinfo{year}{2009}) \bibinfo{pages}{2679--2693}.
%Type = Article
\bibitem[{Iordache et~al.(2012)Iordache, Bioucas-Dias, and Plaza}]{iordache2012total}
\bibinfo{author}{M.-D. Iordache}, \bibinfo{author}{J.~M. Bioucas-Dias}, \bibinfo{author}{A.~Plaza},
\newblock \bibinfo{title}{Total variation spatial regularization for sparse hyperspectral unmixing},
\newblock \bibinfo{journal}{IEEE Transactions on Geoscience and Remote Sensing} \bibinfo{volume}{50} (\bibinfo{year}{2012}) \bibinfo{pages}{4484--4502}.
%Type = Inproceedings
\bibitem[{Bioucas-Dias and Figueiredo(2010)}]{bioucas2010alternating}
\bibinfo{author}{J.~M. Bioucas-Dias}, \bibinfo{author}{M.~A. Figueiredo},
\newblock \bibinfo{title}{Alternating direction algorithms for constrained sparse regression: Application to hyperspectral unmixing},
\newblock in: \bibinfo{booktitle}{2010 2nd Workshop on Hyperspectral Image and Signal Processing: Evolution in Remote Sensing}, \bibinfo{organization}{IEEE}, \bibinfo{year}{2010}, pp. \bibinfo{pages}{1--4}.
%Type = Article
\bibitem[{Iordache et~al.(2013{\natexlab{a}})Iordache, Bioucas-Dias, and Plaza}]{iordache2013collaborative}
\bibinfo{author}{M.-D. Iordache}, \bibinfo{author}{J.~M. Bioucas-Dias}, \bibinfo{author}{A.~Plaza},
\newblock \bibinfo{title}{Collaborative sparse regression for hyperspectral unmixing},
\newblock \bibinfo{journal}{IEEE Transactions on geoscience and remote sensing} \bibinfo{volume}{52} (\bibinfo{year}{2013}{\natexlab{a}}) \bibinfo{pages}{341--354}.
%Type = Article
\bibitem[{Iordache et~al.(2013{\natexlab{b}})Iordache, Bioucas-Dias, Plaza, and Somers}]{iordache2013music}
\bibinfo{author}{M.-D. Iordache}, \bibinfo{author}{J.~M. Bioucas-Dias}, \bibinfo{author}{A.~Plaza}, \bibinfo{author}{B.~Somers},
\newblock \bibinfo{title}{Music-csr: Hyperspectral unmixing via multiple signal classification and collaborative sparse regression},
\newblock \bibinfo{journal}{IEEE Transactions on Geoscience and Remote Sensing} \bibinfo{volume}{52} (\bibinfo{year}{2013}{\natexlab{b}}) \bibinfo{pages}{4364--4382}.
%Type = Article
\bibitem[{Heylen et~al.(2010)Heylen, Burazerovic, and Scheunders}]{heylen2010non}
\bibinfo{author}{R.~Heylen}, \bibinfo{author}{D.~Burazerovic}, \bibinfo{author}{P.~Scheunders},
\newblock \bibinfo{title}{Non-linear spectral unmixing by geodesic simplex volume maximization},
\newblock \bibinfo{journal}{IEEE Journal of Selected Topics in Signal Processing} \bibinfo{volume}{5} (\bibinfo{year}{2010}) \bibinfo{pages}{534--542}.
%Type = Article
\bibitem[{Shkuratov et~al.(1999)Shkuratov, Starukhina, Hoffmann, and Arnold}]{shkuratov1999model}
\bibinfo{author}{Y.~Shkuratov}, \bibinfo{author}{L.~Starukhina}, \bibinfo{author}{H.~Hoffmann}, \bibinfo{author}{G.~Arnold},
\newblock \bibinfo{title}{A model of spectral albedo of particulate surfaces: Implications for optical properties of the moon},
\newblock \bibinfo{journal}{Icarus} \bibinfo{volume}{137} (\bibinfo{year}{1999}) \bibinfo{pages}{235--246}.
%Type = Inproceedings
\bibitem[{Broadwater and Banerjee(2009)}]{broadwater2009comparison}
\bibinfo{author}{J.~Broadwater}, \bibinfo{author}{A.~Banerjee},
\newblock \bibinfo{title}{A comparison of kernel functions for intimate mixture models},
\newblock in: \bibinfo{booktitle}{2009 First Workshop on Hyperspectral Image and Signal Processing: Evolution in Remote Sensing}, \bibinfo{organization}{IEEE}, \bibinfo{year}{2009}, pp. \bibinfo{pages}{1--4}.
%Type = Inproceedings
\bibitem[{Broadwater et~al.(2007)Broadwater, Chellappa, Banerjee, and Burlina}]{broadwater2007kernel}
\bibinfo{author}{J.~Broadwater}, \bibinfo{author}{R.~Chellappa}, \bibinfo{author}{A.~Banerjee}, \bibinfo{author}{P.~Burlina},
\newblock \bibinfo{title}{Kernel fully constrained least squares abundance estimates},
\newblock in: \bibinfo{booktitle}{2007 IEEE international geoscience and remote sensing symposium}, \bibinfo{organization}{IEEE}, \bibinfo{year}{2007}, pp. \bibinfo{pages}{4041--4044}.
%Type = Inproceedings
\bibitem[{Broadwater and Banerjee(2011)}]{broadwater2011mapping}
\bibinfo{author}{J.~Broadwater}, \bibinfo{author}{A.~Banerjee},
\newblock \bibinfo{title}{Mapping intimate mixtures using an adaptive kernel-based technique},
\newblock in: \bibinfo{booktitle}{2011 3rd Workshop on Hyperspectral Image and Signal Processing: Evolution in Remote Sensing (WHISPERS)}, \bibinfo{organization}{IEEE}, \bibinfo{year}{2011}, pp. \bibinfo{pages}{1--4}.
%Type = Article
\bibitem[{Chen et~al.(2012)Chen, Richard, and Honeine}]{chen2012nonlinear}
\bibinfo{author}{J.~Chen}, \bibinfo{author}{C.~Richard}, \bibinfo{author}{P.~Honeine},
\newblock \bibinfo{title}{Nonlinear unmixing of hyperspectral data based on a linear-mixture/nonlinear-fluctuation model},
\newblock \bibinfo{journal}{IEEE Transactions on Signal Processing} \bibinfo{volume}{61} (\bibinfo{year}{2012}) \bibinfo{pages}{480--492}.
%Type = Article
\bibitem[{Hong et~al.(2018)Hong, Yokoya, Chanussot, and Zhu}]{hong2018augmented}
\bibinfo{author}{D.~Hong}, \bibinfo{author}{N.~Yokoya}, \bibinfo{author}{J.~Chanussot}, \bibinfo{author}{X.~X. Zhu},
\newblock \bibinfo{title}{An augmented linear mixing model to address spectral variability for hyperspectral unmixing},
\newblock \bibinfo{journal}{IEEE Transactions on Image Processing} \bibinfo{volume}{28} (\bibinfo{year}{2018}) \bibinfo{pages}{1923--1938}.
%Type = Article
\bibitem[{Borsoi et~al.(2021)Borsoi, Imbiriba, Bermudez, Richard, Chanussot, Drumetz, Tourneret, Zare, and Jutten}]{borsoi2021spectral}
\bibinfo{author}{R.~A. Borsoi}, \bibinfo{author}{T.~Imbiriba}, \bibinfo{author}{J.~C.~M. Bermudez}, \bibinfo{author}{C.~Richard}, \bibinfo{author}{J.~Chanussot}, \bibinfo{author}{L.~Drumetz}, \bibinfo{author}{J.-Y. Tourneret}, \bibinfo{author}{A.~Zare}, \bibinfo{author}{C.~Jutten},
\newblock \bibinfo{title}{Spectral variability in hyperspectral data unmixing: A comprehensive review},
\newblock \bibinfo{journal}{IEEE geoscience and remote sensing magazine} \bibinfo{volume}{9} (\bibinfo{year}{2021}) \bibinfo{pages}{223--270}.
%Type = Article
\bibitem[{Drumetz et~al.(2016)Drumetz, Veganzones, Henrot, Phlypo, Chanussot, and Jutten}]{drumetz2016blind}
\bibinfo{author}{L.~Drumetz}, \bibinfo{author}{M.-A. Veganzones}, \bibinfo{author}{S.~Henrot}, \bibinfo{author}{R.~Phlypo}, \bibinfo{author}{J.~Chanussot}, \bibinfo{author}{C.~Jutten},
\newblock \bibinfo{title}{Blind hyperspectral unmixing using an extended linear mixing model to address spectral variability},
\newblock \bibinfo{journal}{IEEE Transactions on Image Processing} \bibinfo{volume}{25} (\bibinfo{year}{2016}) \bibinfo{pages}{3890--3905}.
%Type = Article
\bibitem[{Palsson et~al.(2018)Palsson, Sigurdsson, Sveinsson, and Ulfarsson}]{palsson2018hyperspectral}
\bibinfo{author}{B.~Palsson}, \bibinfo{author}{J.~Sigurdsson}, \bibinfo{author}{J.~R. Sveinsson}, \bibinfo{author}{M.~O. Ulfarsson},
\newblock \bibinfo{title}{Hyperspectral unmixing using a neural network autoencoder},
\newblock \bibinfo{journal}{IEEE Access} \bibinfo{volume}{6} (\bibinfo{year}{2018}) \bibinfo{pages}{25646--25656}.
%Type = Article
\bibitem[{Su et~al.(2019)Su, Li, Plaza, Marinoni, Gamba, and Chakravortty}]{su2019daen}
\bibinfo{author}{Y.~Su}, \bibinfo{author}{J.~Li}, \bibinfo{author}{A.~Plaza}, \bibinfo{author}{A.~Marinoni}, \bibinfo{author}{P.~Gamba}, \bibinfo{author}{S.~Chakravortty},
\newblock \bibinfo{title}{Daen: Deep autoencoder networks for hyperspectral unmixing},
\newblock \bibinfo{journal}{IEEE Transactions on Geoscience and Remote Sensing} \bibinfo{volume}{57} (\bibinfo{year}{2019}) \bibinfo{pages}{4309--4321}.
%Type = Article
\bibitem[{Zhang et~al.(2018)Zhang, Sun, Zhang, Wu, and Jiao}]{zhang2018hyperspectral}
\bibinfo{author}{X.~Zhang}, \bibinfo{author}{Y.~Sun}, \bibinfo{author}{J.~Zhang}, \bibinfo{author}{P.~Wu}, \bibinfo{author}{L.~Jiao},
\newblock \bibinfo{title}{Hyperspectral unmixing via deep convolutional neural networks},
\newblock \bibinfo{journal}{IEEE Geoscience and Remote Sensing Letters} \bibinfo{volume}{15} (\bibinfo{year}{2018}) \bibinfo{pages}{1755--1759}.
%Type = Article
\bibitem[{Li et~al.(2024{\natexlab{a}})Li, Zhang, Hong, Zhou, Vivone, Li, and Chanussot}]{cinese_casformer}
\bibinfo{author}{C.~Li}, \bibinfo{author}{B.~Zhang}, \bibinfo{author}{D.~Hong}, \bibinfo{author}{J.~Zhou}, \bibinfo{author}{G.~Vivone}, \bibinfo{author}{S.~Li}, \bibinfo{author}{J.~Chanussot},
\newblock \bibinfo{title}{Casformer: Cascaded transformers for fusion-aware computational hyperspectral imaging},
\newblock \bibinfo{journal}{Information Fusion} \bibinfo{volume}{108} (\bibinfo{year}{2024}{\natexlab{a}}) \bibinfo{pages}{102408}. \URLprefix \url{https://www.sciencedirect.com/science/article/pii/S1566253524001866}. \DOIprefix\doi{https://doi.org/10.1016/j.inffus.2024.102408}.
%Type = Article
\bibitem[{Li et~al.(2024{\natexlab{b}})Li, Zhang, Hong, Jia, Plaza, and Chanussot}]{cinese_disentangledpriors}
\bibinfo{author}{C.~Li}, \bibinfo{author}{B.~Zhang}, \bibinfo{author}{D.~Hong}, \bibinfo{author}{X.~Jia}, \bibinfo{author}{A.~Plaza}, \bibinfo{author}{J.~Chanussot},
\newblock \bibinfo{title}{Learning disentangled priors for hyperspectral anomaly detection: A coupling model-driven and data-driven paradigm},
\newblock \bibinfo{journal}{IEEE Transactions on Neural Networks and Learning Systems} \bibinfo{volume}{36} (\bibinfo{year}{2024}{\natexlab{b}}) \bibinfo{pages}{6883--6896}.
%Type = Article
\bibitem[{Li et~al.(2025)Li, Hong, Zhang, Li, Camps-Valls, Zhu, and Chanussot}]{cinese_urban}
\bibinfo{author}{C.~Li}, \bibinfo{author}{D.~Hong}, \bibinfo{author}{B.~Zhang}, \bibinfo{author}{Y.~Li}, \bibinfo{author}{G.~Camps-Valls}, \bibinfo{author}{X.~X. Zhu}, \bibinfo{author}{J.~Chanussot},
\newblock \bibinfo{title}{Urbansam: Learning invariance-inspired adapters for segment anything models in urban construction},
\newblock \bibinfo{journal}{arXiv preprint arXiv:2502.15199}  (\bibinfo{year}{2025}).
%Type = Article
\bibitem[{Li et~al.(2019)Li, Song, Fang, Chen, Ghamisi, and Benediktsson}]{cinese_paraculo_review}
\bibinfo{author}{S.~Li}, \bibinfo{author}{W.~Song}, \bibinfo{author}{L.~Fang}, \bibinfo{author}{Y.~Chen}, \bibinfo{author}{P.~Ghamisi}, \bibinfo{author}{J.~A. Benediktsson},
\newblock \bibinfo{title}{Deep learning for hyperspectral image classification: An overview},
\newblock \bibinfo{journal}{IEEE Transactions on Geoscience and Remote Sensing} \bibinfo{volume}{57} (\bibinfo{year}{2019}) \bibinfo{pages}{6690--6709}. \DOIprefix\doi{10.1109/TGRS.2019.2907932}.
%Type = Article
\bibitem[{Chen et~al.(2024)Chen, Chen, Liu, Li, Zou, and Shi}]{cinese_paraculo_mamba}
\bibinfo{author}{K.~Chen}, \bibinfo{author}{B.~Chen}, \bibinfo{author}{C.~Liu}, \bibinfo{author}{W.~Li}, \bibinfo{author}{Z.~Zou}, \bibinfo{author}{Z.~Shi},
\newblock \bibinfo{title}{Rsmamba: Remote sensing image classification with state space model},
\newblock \bibinfo{journal}{IEEE Geoscience and Remote Sensing Letters} \bibinfo{volume}{21} (\bibinfo{year}{2024}) \bibinfo{pages}{1--5}. \DOIprefix\doi{10.1109/LGRS.2024.3407111}.
%Type = Book
\bibitem[{Cressie(2015)}]{cressie2015statistics}
\bibinfo{author}{N.~Cressie}, \bibinfo{title}{Statistics for spatial data}, \bibinfo{publisher}{John Wiley \& Sons}, \bibinfo{year}{2015}.
%Type = Article
\bibitem[{Xie et~al.(2022)Xie, Wu, Mao, Zhou, and Liu}]{xie2022open}
\bibinfo{author}{B.~Xie}, \bibinfo{author}{L.~Wu}, \bibinfo{author}{W.~Mao}, \bibinfo{author}{S.~Zhou}, \bibinfo{author}{S.~Liu},
\newblock \bibinfo{title}{An open integrated rock spectral library (rocksl) for a global sharing and matching service},
\newblock \bibinfo{journal}{Minerals} \bibinfo{volume}{12} (\bibinfo{year}{2022}) \bibinfo{pages}{118}.
%Type = Article
\bibitem[{Kokaly et~al.(2017)Kokaly, Clark, Swayze, Livo, Hoefen, Pearson, Wise, Benzel, Lowers, Driscoll et~al.}]{kokaly2017usgs}
\bibinfo{author}{R.~Kokaly}, \bibinfo{author}{R.~Clark}, \bibinfo{author}{G.~Swayze}, \bibinfo{author}{K.~Livo}, \bibinfo{author}{T.~Hoefen}, \bibinfo{author}{N.~Pearson}, \bibinfo{author}{R.~Wise}, \bibinfo{author}{W.~Benzel}, \bibinfo{author}{H.~Lowers}, \bibinfo{author}{R.~Driscoll}, et~al.,
\newblock \bibinfo{title}{Usgs spectral library version 7 data: Us geological survey data release [dataset]},
\newblock \bibinfo{journal}{United States Geological Survey (USGS): Reston, VA, USA}  (\bibinfo{year}{2017}).
%Type = Article
\bibitem[{Baldridge et~al.(2009)Baldridge, Hook, Grove, and Rivera}]{baldridge2009aster}
\bibinfo{author}{A.~M. Baldridge}, \bibinfo{author}{S.~J. Hook}, \bibinfo{author}{C.~Grove}, \bibinfo{author}{G.~Rivera},
\newblock \bibinfo{title}{The aster spectral library version 2.0},
\newblock \bibinfo{journal}{Remote Sensing of Environment} \bibinfo{volume}{113} (\bibinfo{year}{2009}) \bibinfo{pages}{711--715}.
%Type = Article
\bibitem[{Meerdink et~al.(2019)Meerdink, Hook, Roberts, and Abbott}]{meerdink2019ecostress}
\bibinfo{author}{S.~K. Meerdink}, \bibinfo{author}{S.~J. Hook}, \bibinfo{author}{D.~A. Roberts}, \bibinfo{author}{E.~A. Abbott},
\newblock \bibinfo{title}{The ecostress spectral library version 1.0},
\newblock \bibinfo{journal}{Remote Sensing of Environment} \bibinfo{volume}{230} (\bibinfo{year}{2019}) \bibinfo{pages}{111196}.
%Type = Article
\bibitem[{Christensen et~al.(2000)Christensen, Bandfield, Hamilton, Howard, Lane, Piatek, Ruff, and Stefanov}]{christensen2000thermal}
\bibinfo{author}{P.~R. Christensen}, \bibinfo{author}{J.~L. Bandfield}, \bibinfo{author}{V.~E. Hamilton}, \bibinfo{author}{D.~A. Howard}, \bibinfo{author}{M.~D. Lane}, \bibinfo{author}{J.~L. Piatek}, \bibinfo{author}{S.~W. Ruff}, \bibinfo{author}{W.~L. Stefanov},
\newblock \bibinfo{title}{A thermal emission spectral library of rock-forming minerals},
\newblock \bibinfo{journal}{Journal of Geophysical Research: Planets} \bibinfo{volume}{105} (\bibinfo{year}{2000}) \bibinfo{pages}{9735--9739}.
%Type = Article
\bibitem[{Maturilli et~al.(2008)Maturilli, Helbert, and Moroz}]{maturilli2008berlin}
\bibinfo{author}{A.~Maturilli}, \bibinfo{author}{J.~Helbert}, \bibinfo{author}{L.~Moroz},
\newblock \bibinfo{title}{The berlin emissivity database (bed)},
\newblock \bibinfo{journal}{Planetary and Space Science} \bibinfo{volume}{56} (\bibinfo{year}{2008}) \bibinfo{pages}{420--425}.
%Type = Article
\bibitem[{Loizeau et~al.(2020)Loizeau, Lequertier, Poulet, Hamm, Pilorget, Meslier-Lourit, Lantz, Werner, Rull, and Bibring}]{loizeau2020planetary}
\bibinfo{author}{D.~Loizeau}, \bibinfo{author}{G.~Lequertier}, \bibinfo{author}{F.~Poulet}, \bibinfo{author}{V.~Hamm}, \bibinfo{author}{C.~Pilorget}, \bibinfo{author}{L.~Meslier-Lourit}, \bibinfo{author}{C.~Lantz}, \bibinfo{author}{S.~C. Werner}, \bibinfo{author}{F.~Rull}, \bibinfo{author}{J.-P. Bibring},
\newblock \bibinfo{title}{Planetary terrestrial analogues library project: 2. building a laboratory facility for micromega characterization},
\newblock \bibinfo{journal}{Planetary and Space Science} \bibinfo{volume}{193} (\bibinfo{year}{2020}) \bibinfo{pages}{105087}.
%Type = Article
\bibitem[{Percival et~al.(2018)Percival, Bosman, Potter, Peter, Laudadio, Abraham, Shiley, and Sherry}]{percival2018customized}
\bibinfo{author}{J.~B. Percival}, \bibinfo{author}{S.~A. Bosman}, \bibinfo{author}{E.~G. Potter}, \bibinfo{author}{J.~M. Peter}, \bibinfo{author}{A.~B. Laudadio}, \bibinfo{author}{A.~C. Abraham}, \bibinfo{author}{D.~A. Shiley}, \bibinfo{author}{C.~Sherry},
\newblock \bibinfo{title}{Customized spectral libraries for effective mineral exploration: mining national mineral collections},
\newblock \bibinfo{journal}{Clays and Clay Minerals} \bibinfo{volume}{66} (\bibinfo{year}{2018}) \bibinfo{pages}{297--314}.
%Type = Misc
\bibitem[{{NASA Planetary Data System}(2024)}]{pds-dataset}
\bibinfo{author}{{NASA Planetary Data System}}, \bibinfo{title}{Pds geosciences node spectral library [dataset]}, \bibinfo{howpublished}{https://pds-geosciences.wustl.edu/spectrallibrary/default.htm}, \bibinfo{year}{2024}.
%Type = Misc
\bibitem[{Swayze et~al.(1992)Swayze, Clark, Kruse, Sutley, and Gallagher}]{swayze1992ground}
\bibinfo{author}{G.~Swayze}, \bibinfo{author}{R.~N. Clark}, \bibinfo{author}{F.~Kruse}, \bibinfo{author}{S.~Sutley}, \bibinfo{author}{A.~Gallagher}, \bibinfo{title}{Ground-truthing aviris mineral mapping at cuprite, nevada}, \bibinfo{year}{1992}.
%Type = Article
\bibitem[{Swayze et~al.(2014)Swayze, Clark, Goetz, Livo, Breit, Kruse, Sutley, Snee, Lowers, Post et~al.}]{swayze2014mapping}
\bibinfo{author}{G.~A. Swayze}, \bibinfo{author}{R.~N. Clark}, \bibinfo{author}{A.~F. Goetz}, \bibinfo{author}{K.~E. Livo}, \bibinfo{author}{G.~N. Breit}, \bibinfo{author}{F.~A. Kruse}, \bibinfo{author}{S.~J. Sutley}, \bibinfo{author}{L.~W. Snee}, \bibinfo{author}{H.~A. Lowers}, \bibinfo{author}{J.~L. Post}, et~al.,
\newblock \bibinfo{title}{Mapping advanced argillic alteration at cuprite, nevada, using imaging spectroscopy},
\newblock \bibinfo{journal}{Economic geology} \bibinfo{volume}{109} (\bibinfo{year}{2014}) \bibinfo{pages}{1179--1221}.
%Type = Article
\bibitem[{Clark et~al.(2003)Clark, Swayze, Livo, Kokaly, Sutley, Dalton, McDougal, and Gent}]{clark2003imaging}
\bibinfo{author}{R.~N. Clark}, \bibinfo{author}{G.~A. Swayze}, \bibinfo{author}{K.~E. Livo}, \bibinfo{author}{R.~F. Kokaly}, \bibinfo{author}{S.~J. Sutley}, \bibinfo{author}{J.~B. Dalton}, \bibinfo{author}{R.~R. McDougal}, \bibinfo{author}{C.~A. Gent},
\newblock \bibinfo{title}{Imaging spectroscopy: Earth and planetary remote sensing with the usgs tetracorder and expert systems},
\newblock \bibinfo{journal}{Journal of Geophysical Research: Planets} \bibinfo{volume}{108} (\bibinfo{year}{2003}).
%Type = Misc
\bibitem[{{MathWorks Image Processing Team}(2024)}]{matlab}
\bibinfo{author}{{MathWorks Image Processing Team}}, \bibinfo{title}{Hyperspectral imaging library for image processing toolbox}, \bibinfo{howpublished}{https://www.mathworks.com/matlabcentral/fileexchange/76796-hyperspectral-imaging-library-for-image-processing-toolbox}, \bibinfo{year}{2024}.
%Type = Article
\bibitem[{Nascimento and Dias(2005)}]{nascimento2005does}
\bibinfo{author}{J.~M. Nascimento}, \bibinfo{author}{J.~M. Dias},
\newblock \bibinfo{title}{Does independent component analysis play a role in unmixing hyperspectral data?},
\newblock \bibinfo{journal}{IEEE Transactions on Geoscience and Remote Sensing} \bibinfo{volume}{43} (\bibinfo{year}{2005}) \bibinfo{pages}{175--187}.
%Type = Article
\bibitem[{Heinz et~al.(2001)}]{heinz2001fully}
\bibinfo{author}{D.~C. Heinz}, et~al.,
\newblock \bibinfo{title}{Fully constrained least squares linear spectral mixture analysis method for material quantification in hyperspectral imagery},
\newblock \bibinfo{journal}{IEEE transactions on geoscience and remote sensing} \bibinfo{volume}{39} (\bibinfo{year}{2001}) \bibinfo{pages}{529--545}.
%Type = Article
\bibitem[{Murchie et~al.(2007)Murchie, Arvidson, Bedini, Beisser, Bibring, Bishop, Boldt, Cavender, Choo, Clancy et~al.}]{murchie2007compact}
\bibinfo{author}{S.~Murchie}, \bibinfo{author}{R.~Arvidson}, \bibinfo{author}{P.~Bedini}, \bibinfo{author}{K.~Beisser}, \bibinfo{author}{J.-P. Bibring}, \bibinfo{author}{J.~Bishop}, \bibinfo{author}{J.~Boldt}, \bibinfo{author}{P.~Cavender}, \bibinfo{author}{T.~Choo}, \bibinfo{author}{R.~Clancy}, et~al.,
\newblock \bibinfo{title}{Compact reconnaissance imaging spectrometer for mars (crism) on mars reconnaissance orbiter (mro)},
\newblock \bibinfo{journal}{Journal of Geophysical Research: Planets} \bibinfo{volume}{112} (\bibinfo{year}{2007}).
%Type = Article
\bibitem[{Ceamanos et~al.(2011)Ceamanos, Dout{\'e}, Luo, Schmidt, Jouannic, and Chanussot}]{ceamanos2011intercomparison}
\bibinfo{author}{X.~Ceamanos}, \bibinfo{author}{S.~Dout{\'e}}, \bibinfo{author}{B.~Luo}, \bibinfo{author}{F.~Schmidt}, \bibinfo{author}{G.~Jouannic}, \bibinfo{author}{J.~Chanussot},
\newblock \bibinfo{title}{Intercomparison and validation of techniques for spectral unmixing of hyperspectral images: A planetary case study},
\newblock \bibinfo{journal}{IEEE Transactions on Geoscience and Remote Sensing} \bibinfo{volume}{49} (\bibinfo{year}{2011}) \bibinfo{pages}{4341--4358}.
%Type = Inproceedings
\bibitem[{Bibring et~al.(2004)Bibring, Soufflot, Berth{\'e}, Langevin, Gondet, Drossart, Bouy{\'e}, Combes, Puget, Semery et~al.}]{bibring2004omega}
\bibinfo{author}{J.-P. Bibring}, \bibinfo{author}{A.~Soufflot}, \bibinfo{author}{M.~Berth{\'e}}, \bibinfo{author}{Y.~Langevin}, \bibinfo{author}{B.~Gondet}, \bibinfo{author}{P.~Drossart}, \bibinfo{author}{M.~Bouy{\'e}}, \bibinfo{author}{M.~Combes}, \bibinfo{author}{P.~Puget}, \bibinfo{author}{A.~Semery}, et~al.,
\newblock \bibinfo{title}{Omega: Observatoire pour la min{\'e}ralogie, l'eau, les glaces et l'activit{\'e}},
\newblock in: \bibinfo{booktitle}{Mars Express: the scientific payload}, volume \bibinfo{volume}{1240}, \bibinfo{year}{2004}, pp. \bibinfo{pages}{37--49}.
%Type = Article
\bibitem[{Liu et~al.(2018)Liu, Luo, Dout{\'e}, and Chanussot}]{liu2018exploration}
\bibinfo{author}{J.~Liu}, \bibinfo{author}{B.~Luo}, \bibinfo{author}{S.~Dout{\'e}}, \bibinfo{author}{J.~Chanussot},
\newblock \bibinfo{title}{Exploration of planetary hyperspectral images with unsupervised spectral unmixing: A case study of planet mars},
\newblock \bibinfo{journal}{Remote Sensing} \bibinfo{volume}{10} (\bibinfo{year}{2018}) \bibinfo{pages}{737}.
%Type = Article
\bibitem[{Zhu et~al.(2014)Zhu, Wang, Xiang, Fan, and Pan}]{zhu2014structured}
\bibinfo{author}{F.~Zhu}, \bibinfo{author}{Y.~Wang}, \bibinfo{author}{S.~Xiang}, \bibinfo{author}{B.~Fan}, \bibinfo{author}{C.~Pan},
\newblock \bibinfo{title}{Structured sparse method for hyperspectral unmixing},
\newblock \bibinfo{journal}{ISPRS Journal of Photogrammetry and Remote Sensing} \bibinfo{volume}{88} (\bibinfo{year}{2014}) \bibinfo{pages}{101--118}.
%Type = Article
\bibitem[{White et~al.(2022)White, Frye, Christensen, Gelfand, and Silander}]{gelfand}
\bibinfo{author}{P.~A. White}, \bibinfo{author}{H.~Frye}, \bibinfo{author}{M.~F. Christensen}, \bibinfo{author}{A.~E. Gelfand}, \bibinfo{author}{J.~A. Silander},
\newblock \bibinfo{title}{Spatial functional data modeling of plant reflectances},
\newblock \bibinfo{journal}{The Annals of Applied Statistics} \bibinfo{volume}{16} (\bibinfo{year}{2022}) \bibinfo{pages}{1919--1936}.
%Type = Misc
\bibitem[{Zammit-Mangion et~al.(2024)Zammit-Mangion, Kaminski, Tran, Filippone, and Cressie}]{cressiespatialbayesianneuralnetworks}
\bibinfo{author}{A.~Zammit-Mangion}, \bibinfo{author}{M.~D. Kaminski}, \bibinfo{author}{B.-H. Tran}, \bibinfo{author}{M.~Filippone}, \bibinfo{author}{N.~Cressie}, \bibinfo{title}{Spatial bayesian neural networks}, \bibinfo{year}{2024}. \URLprefix \url{https://arxiv.org/abs/2311.09491}. \href{http://arxiv.org/abs/2311.09491}{{\tt arXiv:2311.09491}}.

\end{thebibliography}
%%\end{thebibliography}
\end{document}